\documentclass[11pt,a4paper]{article}
\pdfoutput=1

\usepackage{graphicx}
\usepackage{bmpsize}
\usepackage{jheppub}
\usepackage{comment}

\newcommand{\fulltoday}{\number\day\space \ifcase\month\or
    January\or February\or March\or April\or May\or June\or
    July\or August\or September\or October\or November\or December\fi
    \space\number\year}

 \makeatletter
    
    \@addtoreset{equation}{section}
  \makeatother

\allowdisplaybreaks[3]

\title{\boldmath 
Big Bang Nucleosynthesis Hunts Chameleon Dark Matter}

\author[a]{Hua Chen,}
\author[a]{Taishi Katsuragawa,}
\author[b]{Shinya Matsuzaki}
\author[a]{and Taotao Qiu}

\affiliation[a]{Institute of Astrophysics, Central China Normal University,\\ No. 152 Luoyu Road, Wuhan 430079, China}
\affiliation[b]{Center for Theoretical Physics and College of Physics, Jilin University,\\ Qianjin Street No. 2699, Changchun, 130012, China}

\emailAdd{huachen@mails.ccnu.edu.cn}
\emailAdd{taishi@mail.ccnu.edu.cn}
\emailAdd{synya@jlu.edu.cn}
\emailAdd{qiutt@mail.ccnu.edu.cn}

\abstract{
We study the chameleon field dark matter, dubbed \textit{scalaron}, in $F(R)$ gravity in the Big Bang Nucleosynthesis (BBN) epoch. 
With an $R^{2}$-correction term required to solve the singularity problem for $F(R)$ gravity,
we first find that the scalaron dynamics is governed by the $R^{2}$ term and the chameleon mechanism in the early universe,
which makes the scalaron physics model-independent regarding the low-energy scale modification.
In viable $F(R)$ dark energy models including the $R^{2}$ correction, 
our analysis suggests the scalaron universally evolves in a way with a
bouncing oscillation
irrespective of the low-energy modification for the late-time cosmic acceleration.
Consequently, we find a universal bound on the scalaron mass in the BBN epoch, 
to be reflected on the constraint for the coupling strength of the $R^2$ term, 
which turns out to be more stringent than the one coming from the fifth force experiments. 
It is then shown that 
the scalaron naturally develops a small enough fluctuation 
in the BBN epoch, hence can avoid the current BBN constraint placed by the latest 
Planck 2018 data, and can also have a large enough sensitivity to be 
hunted by the BBN, with more accurate measurements for light element abundances as well 
as the baryon number density fraction. 
}

\begin{document}
\maketitle
\flushbottom


\section{Introduction}
\label{sec1}

While the observations~\cite{Riess:1998cb, Ade:2015xua}, again and again, suggest the existence of dark energy and dark matter, their nature remains unknown: 
what are they and how they arise? 
Such a dark side of the universe strongly indicates extensions of general relativity (GR) or standard model (SM) of particle physics. 
In cosmology, the $\Lambda$CDM model fits the observation very well, although it relies on the fine-tuning of the cosmological constant. 
As an alternative,
a large number of works have attempted to modify Einstein's gravity to dynamically explain the dark energy problem (see~\cite{Nojiri:2017ncd} for a thorough review). 
Among them, $F(R)$ gravity is one of the straightforward extensions for Einstein's gravity.

In order to be realistic, the modified gravities have to recover the GR at small scales or dense regions. 
For example, $F(R)$ gravity fails to meet such a requirement
without the chameleon mechanism~\cite{Khoury:2003aq}. 
With the chameleon mechanism, 
$F(R)$ gravity introduces an additional field degree of freedom in Einstein frame, 
which could serve as dynamical dark energy. 
There are several viable models applied to the late-time cosmology;
for instance, Hu-Sawicki model~\cite{Hu:2007nk}, Starobinsky model~\cite{Starobinsky:2007hu}, Tsujikawa model~\cite{Tsujikawa:2007xu}, and exponential model~\cite{Elizalde:2010ts}.
However, as was revealed in~\cite{Frolov:2008uf, Bamba:2008ut}, dark energy models in $F(R)$ gravity generally suffer from the so-called singularity problems. 
Thus, we need further modifications to cure the singularity problems,
and a resolution by adding a second-order correction $\alpha R^2$ was proposed in \cite{Kobayashi:2008wc, Dev:2008rx}.

As for the dark matter, 
although it is possible to regard the dark matter as a pure gravitational effect in the modified gravity \cite{Milgrom:1983ca}, 
the cosmic microwave background (CMB) data fits well 
with the $\Lambda$CDM model, and also the evidence of rotation curve~\cite{Rubin:1980zd} and weak gravitational lensing~\cite{Clowe:2006eq} 
favor the presence of dark matter as particles. 
Therefore, it is widely believed that 
the existence of dark matter is associated with extensions of the SM 
which could also have some correlation with other mysteries in the universe; for instance, the axion-dark matter that is responsible for the violation of charge and parity (CP) symmetry in 
the quantum chromodynamics (see ~\cite{Marsh:2015xka, Odintsov:2019mlf, Odintsov:2019evb}, for a recent review on the cosmological impact of axion-dark matter). 

On the contrary to the folklore, it has been suggested that
dark matters may also arise from a gravitational origin~\cite{Nojiri:2008nt},  
for instance through an $R^2$ gravity~\cite{Cembranos:2008gj}. 
Inspired by those works,
two of authors recently proposed a scenario to make both the dark energy and dark matter played by a chameleon field, dubbed a scalaron,
in the so-called Starobinsky model of $F(R)$ gravity~\cite{Katsuragawa:2017wge, Katsuragawa:2016yir}.  
There the dark side originates from the chameleon field,
whose potential minimum corresponds to the dark energy and the excitation mode around the potential minimum behaves like the dark matter. 
As a consequence, we can address the coincidence problem~\cite{Katsuragawa:2017wge}. 
A similar scenario has also been applied to the logarithmic $F(R)$ model recently~\cite{Inagaki:2019dqv}.

The scalaron can have other significant characteristics in the sense of astrophysical and phenomenological consequences: 
The scalaron mass becomes massive in the dense region while gets light in the cosmic background.
The scalaron can thus avoid being detected in the current dark matter searches but affects the cosmological (possibly, astrophysical) phenomena as the dark matter, 
which is compared to other scalar-dark matter scenarios, such as axion(-like) particles~\cite{Marsh:2015xka}, 
light scalar dark matter \cite{Boehm:2003hm, Knapen:2017xzo}, 
and ultra-light scalar dark matter~\cite{Lee:2017qve}.  
In further comparison with some dark matter models which also predict a variable mass \cite{GarciaBellido:1992de, Anderson:1997un}, the scalaron could provide a 
characteristic signal.

In this paper, 
we first show that 
with an $R^{2}$-correction term required to solve the singularity problem for $F(R)$ gravity,
the scalaron dynamics is governed by the $R^{2}$ term and the chameleon mechanism in the early universe.
It makes the scalaron physics model-independent regarding the low-energy scale modification in the early universe. 
In viable $F(R)$ dark energy models including the $R^{2}$ correction, 
our analysis suggests the scalaron universally evolves in a way with a bouncing oscillation
irrespective of the low-energy modification for the late-time cosmic acceleration.

In relation to the dark matter physics in the early universe,
the Big Bang Nucleosynthesis (BBN) would give us the stringent observational constraint on the new particles beyond the SM of particle physics.
Therefore, one may naively expect that
the BBN constraint should be applied to the scalaron dark matter, which, 
in contrast to the other dark matter candidate,
has the environment-dependence reflecting the chameleon mechanism.

We find a generic bound on the scalaron mass in the BBN epoch, 
to feedback to the constraint on the coupling strength for the $R^2$ term, 
which is universally applicable to the viable modified gravity.  
It turns out that this bound is more stringent than the one placed by the fifth force experiments. 
We then demonstrate that 
the scalaron naturally develops a small enough fluctuation in the BBN epoch, hence can avoid the current BBN constraint placed by the latest Planck 2018 data. 
The size of the detection sensitivity depends on the initial condition 
for the scalaron fluctuation in the early universe, so 
the BBN data with more accurate measurements for light element abundances as well 
as the baryon number density fraction can 
exclude, or probe some parts of the scalaron scenario.

Throughout the main body of the present paper, 
we will work on the flat Friedmann-Lemaitre-Robertson-Walker 
(FLRW) metric, $\mathrm{diag} \left(-1,\ a^2,\ a^2,\ a^2\right)$, 
while in the appended discussion given in Appendix~\ref{appendix1},
we will take the Minkowski metric with $\mathrm{diag} \left(-1,\ 1,\ 1,\ 1\right)$ for convenience.

\section{Dark energy and eark matter in $F(R)$ gravity}
\label{sec2}

This section provides the brief introduction of the $F(R)$ gravity motivated by the late-time cosmic acceleration.
$F(R)$ models for the dark energy, in general, suffer from the singularity problem arising at a finite value of the field,
which can be improved with the higher curvature term,
and we introduce the model of our interest in this paper.
We also mention the concept and advantage to unify the dark energy and dark matter in $F(R)$ gravity.

\subsection{Dynamical dark energy in $F(R)$ gravity}

In this subsection, we review the $F(R)$ gravity and related dark energy models. 
The action of $F(R)$ gravity reads
\begin{align}
S=\frac{1}{2\kappa^{2}}\int d^{4}x\sqrt{-g}F(R)+\int d^{4}x\sqrt{-g}{\cal L}_\mathrm{Matter}(g^{\mu\nu},\Phi)\,.
\end{align}
Here $\kappa^{2} = 1 / M^{2}_\mathrm{pl}$  
with
$M_\mathrm{pl}$ being the reduced Planck mass, 
$M_\mathrm{pl} \simeq 2\times10^{18}~[\mathrm{GeV}]$.
$\mathcal{L}_\mathrm{Matter}$ denotes the Lagrangian for a matter field $\Phi$,
and the matter field $\Phi$ follows the geodesics of a metric $g_{\mu \nu}$.
According to the Weyl transformation,
\begin{align}
\tilde{g}_{\mu\nu}=\mathrm{e}^{2\sqrt{1/6}\kappa\varphi}g_{\mu\nu}\equiv F_{R}(R)g_{\mu\nu}\,,
\label{eq: Weyl transformation}
\end{align}
we can recover the Einstein's gravity with an additional scalar field $\varphi$, 
so-called the scalaron, in the Einstein frame:
\begin{align}
S&=\frac{1}{2\kappa^{2}}\int d^{4}x\sqrt{-\tilde{g}}\tilde{R}\notag\\
&+\int d^{4}x\sqrt{-\tilde{g}}\left[-\frac{1}{2}\tilde{g}^{\mu\nu}(\partial_{\mu}\varphi)(\partial_{\nu}\varphi)-V(\varphi)\right]\notag\\
&+\int d^{4}x\sqrt{-\tilde{g}}\,\mathrm{e}^{-4\sqrt{1/6}\kappa\varphi}\mathcal{L}_{\mathrm{Matter}}\left[g^{\mu\nu}=\mathrm{e}^{2\sqrt{1/6}\kappa\varphi}\tilde{g}^{\mu\nu},\Phi\right]\,,
\label{Eq: Action in Einstein frame}
\end{align}
where the scalaron potential $V(\varphi)$ is defined as
\begin{align}
V(\varphi)\equiv\frac{1}{2\kappa^{2}}\frac{RF_{R}(R)-F(R)}{F_{R}^{2}(R)}\,.
\label{Original potential}
\end{align}
The variation of the action Eq.~\eqref{Eq: Action in Einstein frame} with respect to $\tilde{g}^{\mu\nu}$ gives
\begin{align}
\tilde{R}_{\mu\nu}-\frac{1}{2}\tilde{R}\tilde{g}_{\mu\nu}-\kappa^{2}\left(-\frac{1}{2}\tilde{g}_{\mu\nu}\tilde{g}^{\rho\sigma}(\partial_{\rho}\varphi)(\partial_{\sigma}\varphi)+(\partial_{\mu}\varphi)(\partial_{\nu}\varphi)-\tilde{g}_{\mu\nu}V(\varphi)\right)&=\kappa^{2}\tilde{T}_{\mu\nu}\,,
\label{Eq: Field equation}
\end{align}
where
\begin{align}
\tilde{T}_{\mu\nu}&\equiv\frac{-2}{\sqrt{-\tilde{g}}}\frac{\delta\left(\sqrt{-\tilde{g}}\,e^{-4\sqrt{1/6}\kappa\varphi}\mathcal{L}_{\mathrm{M}}\left(g^{\mu\nu},\Phi\right)\right)}{\delta\tilde{g}^{\mu\nu}}\notag\\
&=\frac{-2}{e^{4\sqrt{1/6}\kappa\varphi}\sqrt{-g}}\frac{\delta\left(\sqrt{-g}\mathcal{L}_{\mathrm{Matter}}\left(g^{\mu\nu},\Phi\right)\right)}{e^{-2\sqrt{1/6}\kappa\varphi}\delta g^{\mu\nu}}\notag\\
&=e^{-2\sqrt{1/6}\kappa\varphi}T_{\mu\nu}\left(g^{\mu\nu},\Phi\right)\,.
\label{friedman}
\end{align}

The variation of the action Eq.~\eqref{Eq: Action in Einstein frame} with respect to 
the scalaron $\varphi$ gives
\begin{align}
\tilde{\Box}\varphi = V_{,\varphi}(\varphi) + \frac{\kappa}{\sqrt{6}}e^{-4\sqrt{1/6}\kappa\varphi}T^\mu_\mu\, ,
\label{Eq: Equation of motion}
\end{align}
where $\tilde{\Box} = \tilde{\nabla}^{\mu} \tilde{\nabla}_{\mu}$, and $\tilde{\nabla}_{\mu}$ denotes the covariant derivative composed of the Einstein-frame metric $\tilde{g}_{\mu \nu}$.
Then we define the effective potential,
\begin{align}
V_\mathrm{eff}(\varphi)\equiv V(\varphi)  + \int d \varphi\ \frac{\kappa}{\sqrt{6}}e^{-4\sqrt{1/6}\kappa\varphi}T^\mu_\mu \,.
\label{Eq: Effective potential0}
\end{align}
Note that the trace of the energy momentum tensor
$T^{\mu}_{\mu}$ includes a nontrivial dependence of $\varphi$ via the Weyl transformation of the metric in the matter Lagrangian.
When one ignores such a dependence, Eq.~\eqref{Eq: Effective potential0} is reduced to be
\begin{align}
V_\mathrm{eff}(\varphi)\equiv V(\varphi)  - \frac{1}{4}e^{-4\sqrt{1/6}\kappa\varphi}T^\mu_\mu \,,
\label{Eq: Effective potential}
\end{align}
which has been used in the earlier works (for example, \cite{Brax:2004qh}) with a fluid approximation of the matter sector.

It should be noted that one can consider the redefinition of the energy-momentum tensor related to the Weyl transformation. 
The matters defined in the Jordan frame do not satisfy the conservation law defined in the Einstein frame because of the coupling to the scalaron.
Although the redefinition should depend on the type of fundamental field,
proper definitions for the perfect fluids have been already studied (for example, see \cite{Capozziello:2018wul}).
As an illustration, we assume the pressure-less dust where $T^{\mu}_{\ \mu} = -\rho$.
In this case, we have
\begin{align}
\tilde{\Box} \varphi 
= \frac{\partial V_{s}(\varphi)}{\partial \varphi}
- \frac{\kappa}{\sqrt{6}} \rho_{J} \mathrm{e}^{-4 \sqrt{1/6} \kappa \varphi }  
\label{scalaron_eff_potential1}
\, .
\end{align}
Here, the energy density $\rho_{J}$ is defined in the (original) Jordan frame.
On the other hand, it is common to introduce the new energy density $\rho_{E} = \rho \mathrm{e}^{3 \sqrt{1/6} \kappa \varphi}$
which is conserved in the Einstein frame.
Note that the above definition is different from the energy density $\tilde{\rho} = \rho \mathrm{e}^{4 \sqrt{1/6} \kappa \varphi}$
which is consistent with the scaling dimension in the Einstein frame.
Hereafter, we use matter fields originally introduced in the Jordan frame
to take their time-evolutions into account. 
We will discuss the above point later.

The square of the scalaron mass is computed as the second derivative of the effective potential at the potential minimum $\varphi= \varphi_{\min}$:
\begin{align}
m^2 (\varphi_{\min})&\equiv\frac{\partial^2 V_\mathrm{eff}(\varphi_\mathrm{min})}{\partial^2\varphi}\notag\\
&=V(\varphi_{\mathrm{min}})_{,\varphi\varphi}-\frac{2\kappa^2}{3}e^{-4\sqrt{1/6}\kappa\varphi_\mathrm{min}}T^\mu_\mu\, ,
\label{Scalaron mass}
\end{align}
where $\varphi_{\min}$ satisfies $V_{\mathrm{eff} \, , \varphi}(\varphi_{\min}) = 0$.
Due to the exponential coupling between $\varphi$ and $T^\mu_\mu$, 
the effective potential and the mass are environment-dependent quantities. 
Then we can find that there is a positive correlation between the scalaron mass and energy density: 
the scalaron mass becomes heavy in the high-density region, for instance in the solar system, 
while it becomes light in the low-energy region, for instance in the cosmic background. 
Such an environment dependence is called the chameleon mechanism, which enables $F(R)$ gravity to satisfy the observational constraints.

\begin{figure}[!htbp]
\centering
\includegraphics[scale=0.6]{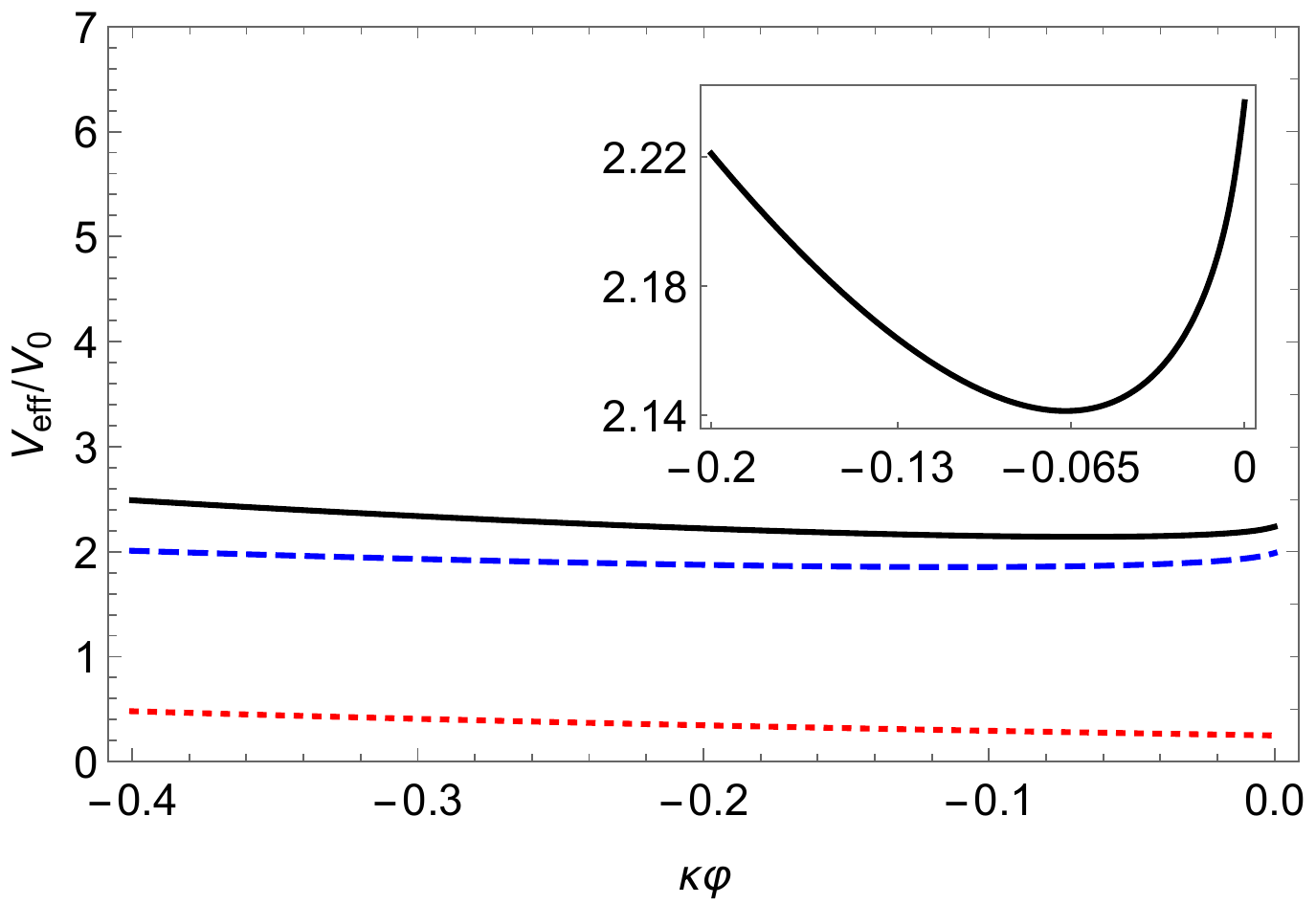}
\caption{The effective potential of Starobinky model with $\beta=2$, $n=1$, and $V_\mathrm{eff}$ being normalized by $V_0=\frac{R_c}{2\kappa^2}$. Here, the dotted red line corresponds to original potential, dashed blue line to matter contribution with $-T^\mu_\mu\sim\rho_\Lambda \sim10^{-47}\ [\mathrm{GeV}^{4}]$ (dark energy density), 
and solid black line to the effective potential.}
\label{Fig: Effective_potential}
\end{figure}

The additional field degree of freedom has been broadly studied as the source of dark energy. 
However, it has been shown that dark energy models in $F(R)$ gravity may suffer from the curvature singularity problem~\cite{Frolov:2008uf}.
To review it in detail, we consider the following dark energy models,
\begin{itemize}
\item[1.] Hu-Sawicki model~\cite{Hu:2007nk}:
\begin{align}
F(R)=R- R_c \frac{c_{1} \left(R/R_c\right)^{n}}{c_{2} \left(R/R_c\right)^{n}+1}\,,
\label{HS}
\end{align}
\item[2.] Starobinsky model~\cite{Starobinsky:2007hu}:
\begin{align}
F(R)=R-\beta R_c \left[1-\left( 1+\frac{R^2}{R^2_c} \right)^{-n} \right]\,,
\label{S}
\end{align}
\item[3.] Tsujikawa model~\cite{Tsujikawa:2007xu}:
\begin{align}
F(R)=R-\beta R_c \tanh\left(\frac{R}{R_c}\right)\,,
\label{T}
\end{align}
\item[4.] Exponential model~\cite{Elizalde:2010ts}:
\begin{align}
F(R)=R-\beta R_c \left( 1 - e^{-R/R_c}\right) \,,
\label{E}
\end{align}
\end{itemize}
where $R_{c}\sim\Lambda\simeq4\times10^{-84}\ [\mathrm{GeV}^{2}]$ corresponds to the dark energy scale, and $c_{i}$, $n$, and $\beta$ are model parameters. 
It is easy to find that $R\rightarrow \infty$ when $\kappa\varphi \rightarrow 0$ in Eq.~\eqref{eq: Weyl transformation}. 
Looking at the Starobinsky model as an example (Fig.~\ref{Fig: Effective_potential}),
we observe that the original potential is lifted due to the matter contribution. 
Eventually, the minimum of the effective potential gets closer to $\kappa\varphi=0$, and the scalaron field can smoothly reach $\kappa\varphi=0$ 
when one considers the perturbation around the minimum, because of the quite shallow potential form. 
This implies the curvature singularity at $R=\infty$ appears at a finite field value and energy level.

To cure the curvature singularity problem above, 
one could extend the models listed in Eqs.~\eqref{HS}, \eqref{S}, \eqref{T}, and \eqref{E}  
by adding a second-order correction, $\alpha R^2$~\cite{Kobayashi:2008wc, Dev:2008rx}. 
Note that those models can explain the dark energy problem in the late-time universe, although there are some differences quantitatively. 
As we will see later, however, 
if we add a second-order correction to cure the singularity problem, 
the differences among those models become negligible in the large-curvature region, and so the models would be indistinguishable in the early universe
when the $R^2$ term dominates. 
So, we shall momentarily work only on the Starobinsky model of our interest.

We should note that $R^2$ correction also plays a significant role in the finite-time future singularity problem~\cite{Bamba:2008ut}.
The earlier works have shown that $R^{2}$ term can remove several types of the above singularities 
although the so-called type IV singularity remains (see Refs.~\cite{Nojiri:2005sx, Capozziello:2009hc} for the classification of these singularities).
As was suggested in~\cite{Nojiri:2008fk}, 
the finite-time future singularity problem has the same origin as the curvature singularity problem stems from,
and thus the curvature singularity can be rephrased as one of the finite-time future singularities.
Furthermore, one can find that not only $R^2$ term but also $R^{n}$, where $1<n<2$, can work to remove those singularities from $F(R)$ gravity.

\subsection{Chameleon field as dark matter in $F(R)$ gravity}

As an illustration, we consider the Starobinsky model with $R^2$ correction.
\begin{align}
F(R)=R-\lambda R_c \left[1-\left( 1+\frac{R^2}{R^2_c} \right)^{-n} \right]+\alpha R^2\,,
\label{Eq: modified Starobinsky model}
\end{align}
where $\alpha \lesssim 10^{22}~\mathrm{[GeV^{-2}]}$ to be consistent with
the constraint from the fifth force experiment~\cite{Katsuragawa:2018wbe}. 
Note that this $R^2$ does not have to be related to an inflaton dynamics in the current analysis.
Since the $R^2$ term will make the curvature sharp as the fluctuating 
$\varphi$ goes to zero, the $\varphi$ cannot perturbatively access the singular point; hence, the singularity problem is to be gone 
if the potential value at $R\rightarrow \infty$ is high enough. 

 Now, we can safely consider the fluctuation of the scalaron field around the potential minimum.
If we quantize such an excitation mode, we obtain the particle picture, which literally brings us a new particle beyond the SM of the particle physics. 
Two of the authors have proposed that this new particle would be a dark matter candidate~\cite{Katsuragawa:2016yir}. 
With this scenario at our hand, 
we may resolve the dark matter problem in the framework of the modified gravity in addition to the dark energy problem, 
which not only gives a unification of the dark side of the universe but also allows us to make the coincidence problem understandable~\cite{Katsuragawa:2017wge}.

Since the $R^2$ term enable us to link the original Starobinsky model to the early universe, 
it is natural to speculate that dark matter is produced in the early universe. 
In relation to the dark matter physics in the early universe,
the BBN gives us the stringent observational constraint on the new particles beyond the SM of particle physics.
Therefore, one may naively expect that 
the BBN constraints should be applied to the scalaron dark matter.
In contrast to the other dark matter candidate, 
the scalaron and, of course, observational constraints on it have the environment-dependence,
reflecting the chameleon mechanism.
In order to address such a {\it chameleon} dark matter,
it is inevitable to introduce an appropriate external matter.
In the following, we shall look into more details on the environment-dependence in the early universe and the BBN constraints on the scalaron.

\section{Chameleon dark matter in early universe}
\label{sec3}

Cosmological studies for the early universe have been investigated with focus on the scalaron dynamics and the chameleon mechanism~\cite{Katsuragawa:2018wbe} (also see \cite{Brax:2004qh} in the scalar-tensor theory). 
In the following analysis, 
we study the time evolution of the scalaron field in the early universe 
during 
the epochs with $T \sim 100$~[GeV] (after the electroweak phase transition) 
down to $\sim 1$~[MeV] (around the BBN), 
by explicitly solving the equation of motion for the scalaron.  
We will adopt a toy-model to evaluate the chameleon mechanism to get a rough prospect for the chameleon in the overall environmental effect through the cosmic history.

\subsection{Analytical approximation}

To help our analytic understanding of the scalaron potential structure, 
by using   
the fluid approximation,  
we shall model the environment surrounding the scalaron overall in the 
target-early universe.  
In this case, the trace of energy-momentum tensor, which shows up in Eqs.~\eqref{Eq: Effective potential} and \eqref{Scalaron mass}, is given by
\begin{align}
T^{\mu}_{\mu} = - (\rho - 3p) \, ,
\end{align}
where $\rho$ and $p$ denote the energy density and pressure of the matters, respectively.
For further convenience,
we define 
\begin{align} 
T^\mu_\mu\equiv-\xi(T)\rho\,, 
\end{align} 
where $\xi(T) = (\rho - 3p) / \rho$ 
and $\xi(T)$ is dimensionless function of temperature $T$. 
We consider the FLRW background, which suffices to study the scalaron evolution in the early universe. 
Then, the $(0,0)$-component of Eq.~\eqref{Eq: Field equation} gives
\begin{align}
3H^{2}=\kappa^{2}\left(\frac{1}{2} \dot{\varphi}^{2} + V(\varphi)
+ e^{-2\sqrt{1/6}\kappa\varphi}\rho_{\mathrm{Matter}}\right)\,.
\label{Eq: Friedmann equation}
\end{align}
Substituting Eq.~\eqref{Eq: Friedmann equation} into the equation of motion, Eq.~\eqref{Eq: Equation of motion}, 
we obtain the time evolution equation for the scalaron, 
\begin{align}
\ddot{\varphi} + 3H\dot{\varphi}+V_{,\varphi}(\varphi)
- \frac{\xi(T)}{\sqrt{6}\kappa}e^{-2\sqrt{1/6}\kappa\varphi}
\left(3H^{2} - \frac{1}{2}\kappa^{2}\dot{\varphi}^{2} - \kappa^{2}V(\varphi) \right) = 0 \,.
\label{Eq: Scalaron_evolution}
\end{align}

Eq.~\eqref{Eq: Scalaron_evolution} is a highly nonlinear differential equation, which is practically so hard to solve. 
To grab a rough insight into the scalaron evolution, 
in this subsection 
we shall first make several approximations to the model, Eq.~\eqref{Eq: modified Starobinsky model},
so that Eq.~\eqref{Eq: Scalaron_evolution} is written in an analytic and manageable form,
which can include the essential part in the early universe.
We will perform the numerical analysis and discuss the validity of those approximations in the next subsection. 

For convenience, we define $r\equiv R/R_c$ and $\gamma\equiv\mathrm{e}^{2\sqrt{1/6}\kappa\varphi}-1$, with $\kappa\varphi=0\Leftrightarrow\gamma=0$. From Eqs.~\eqref{eq: Weyl transformation} and \eqref{Eq: modified Starobinsky model}, we have
\begin{align}
\gamma&=2\alpha R_{c}r-2n\beta\left(1+r^{2}\right)^{-n-1}r\notag\\
&\approx 2\alpha R_{c}r-2n\beta r^{-2n-1}\,,
\end{align} 
where the second line can be achieved if $r \gg 1$ as in our targeted-early universe. 
When $\gamma>0$ ($\kappa\varphi>0$),
\begin{align}
(1+r^{2})^{n+1}&>\frac{n\beta}{\alpha R_{c}}\,,
\end{align}
where $R_c\sim 4\times10^{-84}\ [\mathrm{GeV}^{2}]$, $\beta\gtrsim \mathcal{O}(1)$, $n\gtrsim \mathcal{O}(1)$, and we choose $\alpha \lesssim10^{22}\ [\mathrm{GeV}^{-2}]$ 
consistently with
the constraint from the fifth force experiment~\cite{Katsuragawa:2018wbe}. 

As $r$ is large for our targeted-early universe, 
we can safely neglect the second term in Eq.~\eqref{Eq: modified Starobinsky model}, which corresponds to the present dark energy scale, 
so that the model Eq.~\eqref{Eq: modified Starobinsky model} is approximated as
\begin{align}
F(R)&\approx R+\alpha R^2 \notag\\
&=\frac{1}{2\alpha}\gamma+\frac{1}{4\alpha}\gamma^{2}\,. 
\label{R^2_approx}
\end{align}
Here the second line is achieved by using the Weyl transformation Eq.~\eqref{eq: Weyl transformation}.
It is crucial to note that  
the approximated expression above is also valid for other models quoted in Eqs.~\eqref{HS},~\eqref{T} and~\eqref{E} with the $R^2$ correction, because it is free from the present dark energy scale.

Thus we obtain analytical forms of potential terms as follows:
\begin{align}
V(\varphi)&=\frac{1}{8\kappa^{2} \alpha}\frac{\gamma^{2}}{\left(1+\gamma\right)^{2}}
\label{Simplified potential}
\,, \\
V_{,\varphi}(\varphi)&=\frac{1}{2\alpha\sqrt{6}\kappa}\frac{\gamma}{ \left(1+\gamma\right)^{2}}
\label{First derivative of simplified potential}
\,, \\
V_{,\varphi\varphi}(\varphi)&=\frac{1}{6\alpha}\frac{1-\gamma}{\left(1+\gamma\right)^{2}}
\label{Second derivative of simplified potential}
\,.
\end{align}
The stationary condition for the effective potential  $V_{\mathrm{eff} \, , \varphi}(\varphi_{\min}) = 0$ reads 
\begin{align}
\gamma_{\min} = - 2\kappa^{2}\alpha T^\mu_\mu 
\, ,
\end{align}
where $\gamma_{\min} = \mathrm{e}^{2\sqrt{1/6}\kappa\varphi_{\min}}-1$.
The second derivative of the effective potential is written in terms of $\gamma$ as 
\begin{align}
V_{\mathrm{eff} \, ,\varphi\varphi}(\varphi)
&=\frac{1}{6\alpha}\frac{1-\gamma - 4\kappa^2 \alpha T^\mu_\mu}{\left(1+\gamma\right)^{2}}
\, .
\end{align}
We derive the analytically 
approximated formula for the scalaron mass $m(\varphi_{\min})$ in Eq.~\eqref{Scalaron mass}
\begin{align} 
m(\varphi_{\min}) 
= \sqrt{\frac{1}{6 \alpha}} \cdot \sqrt{\frac{1}{1 - 2 \kappa^2 \alpha T^\mu_\mu}}
\, . 
\label{Scalaron mass-analytic}
\end{align}

\subsection{Numerical evaluation for time evolution of scalaron field}

\begin{figure}[!htbp]
\centering
\includegraphics[scale=0.6]{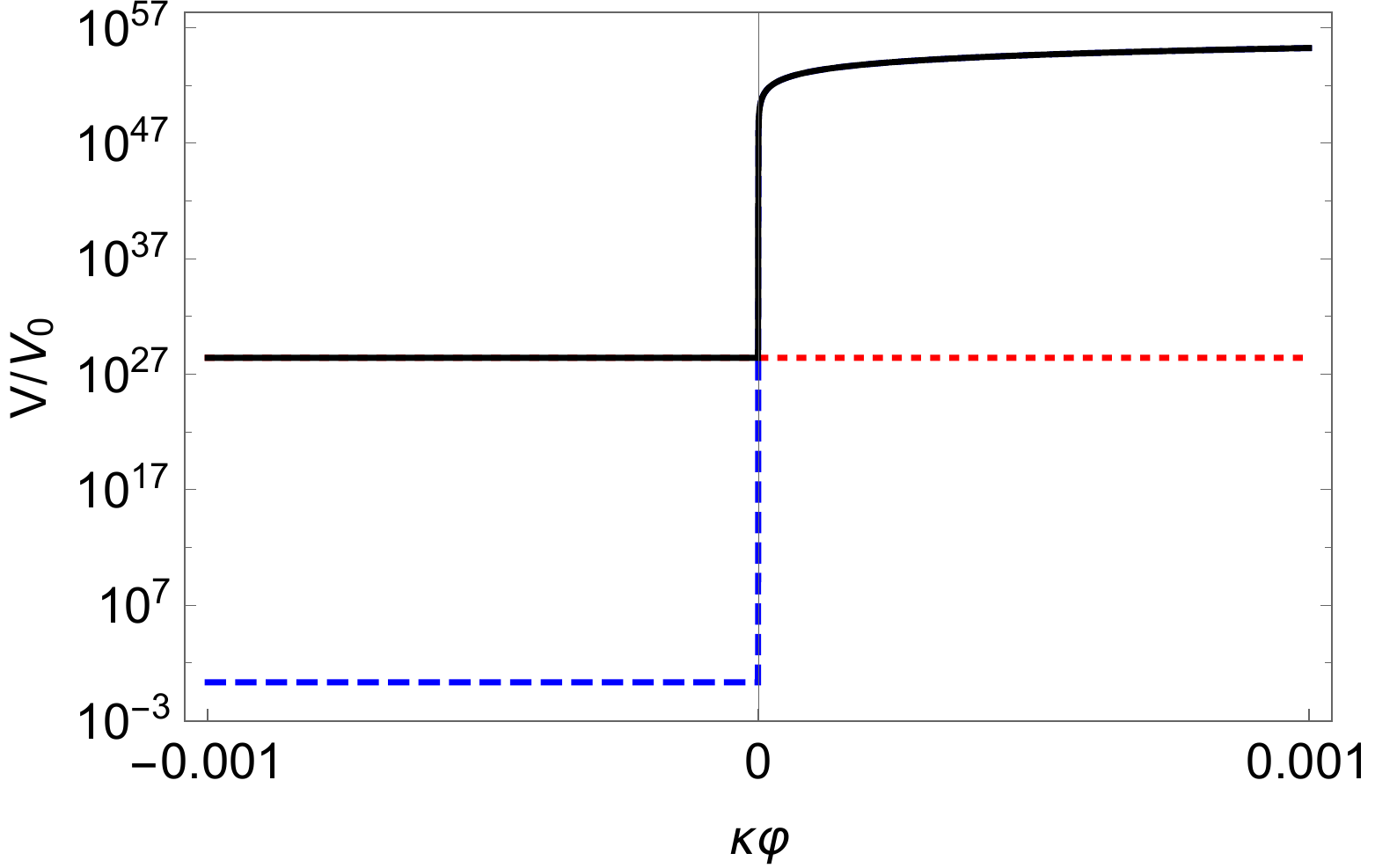}
\caption{The effective potential in the BBN epoch (black solid line), 
which is the sum of the original potential (blue dashed line) and the matter contribution (red dotted line), 
with $-T^\mu_\mu(\mathrm{BBN}) \sim 10^{29}\rho_\Lambda$.}
\label{Fig: Effective_potential_BBN}
\end{figure}

For the numerical analysis, 
we should carefully check whether the above approximations can work or not. 
Firstly, we shall observe the form of the effective potential in Eq.~\eqref{Eq: Effective potential}, 
as depicted in Fig.~\ref{Fig: Effective_potential_BBN},
where we choose 
$\alpha = 10^{22}\ [\mathrm{GeV^{-2}}]$ and $-T^\mu_\mu\sim 10^{-18}\ \mathrm{[GeV^4]}$,
following the previous work~\cite{Katsuragawa:2017wge}.
One could find that in the negative region, $\kappa\varphi<0$,
the matter contribution with $-T^\mu_\mu(\mathrm{BBN})\sim10^{29}\rho_\Lambda$ is highly dominated.
It allows us to safely set the first term to be zero in Eq.~\eqref{Eq: Effective potential}, so that only the matter contribution can be significant in the negative region.

Remarkably enough, 
the potential minimum emerges at around the origin $\varphi=0$,
under the matter contributions in the BBN epoch. 
It happens because of the balance created by the $R^2$ term and nonzero matter contributions, i.e., the chameleon mechanism. 
Thus this fact gives us great support to study the scalaron particle physics in the BBN epoch, as will be described later in details.

\begin{figure}[!htbp]
\centering
\includegraphics[scale=0.6]{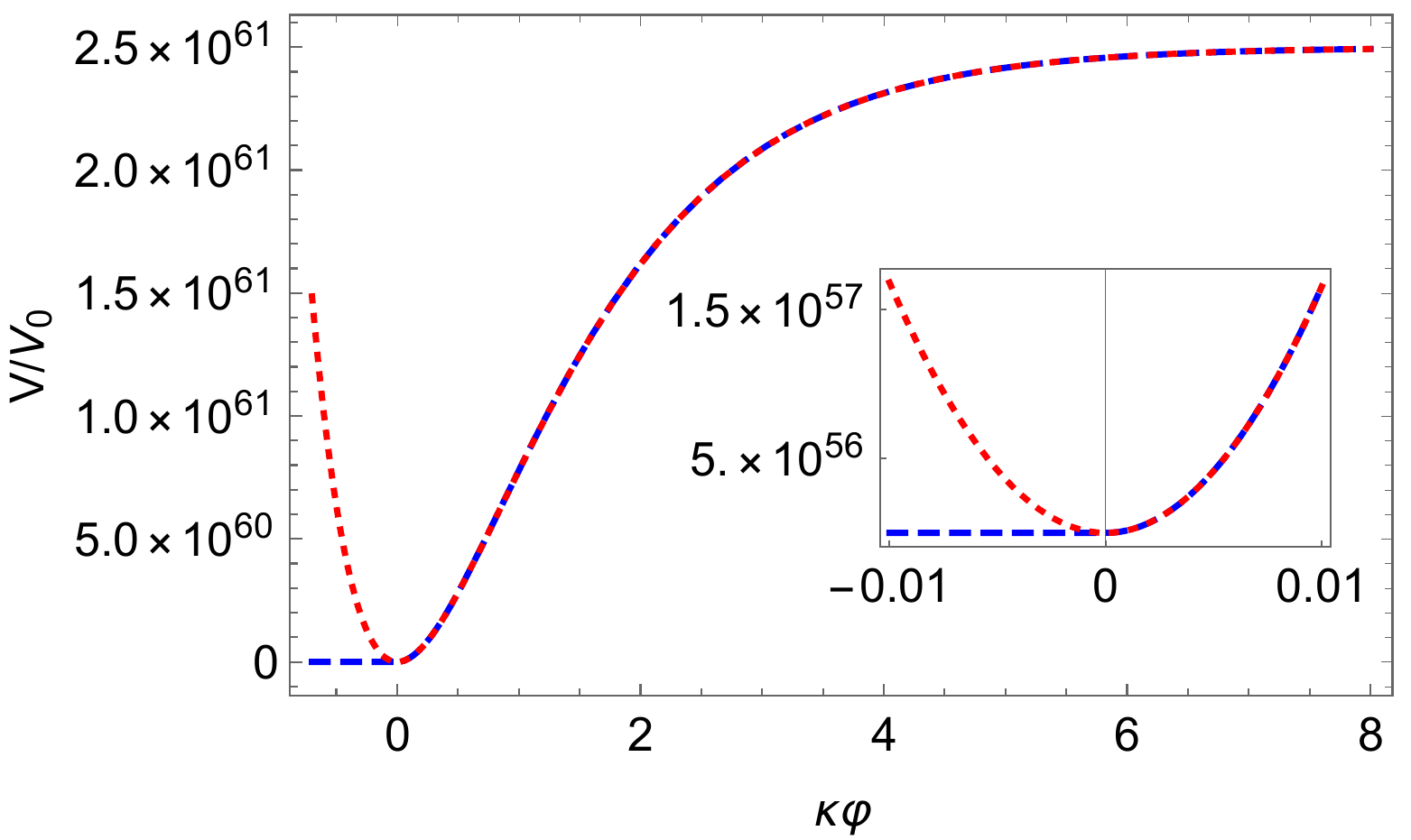}
\caption{The comparison of the approximated potential in Eq.~\eqref{Simplified potential} (red dotted line)
with the original one in Eq.~\eqref{Original potential} (blue dashed line).
The potentials without matter contributions ($\xi=0$) are shown.}
\label{Fig: Potential_original_approximated}
\end{figure}

\begin{figure}[!htbp]
\centering
\includegraphics[scale=0.6]{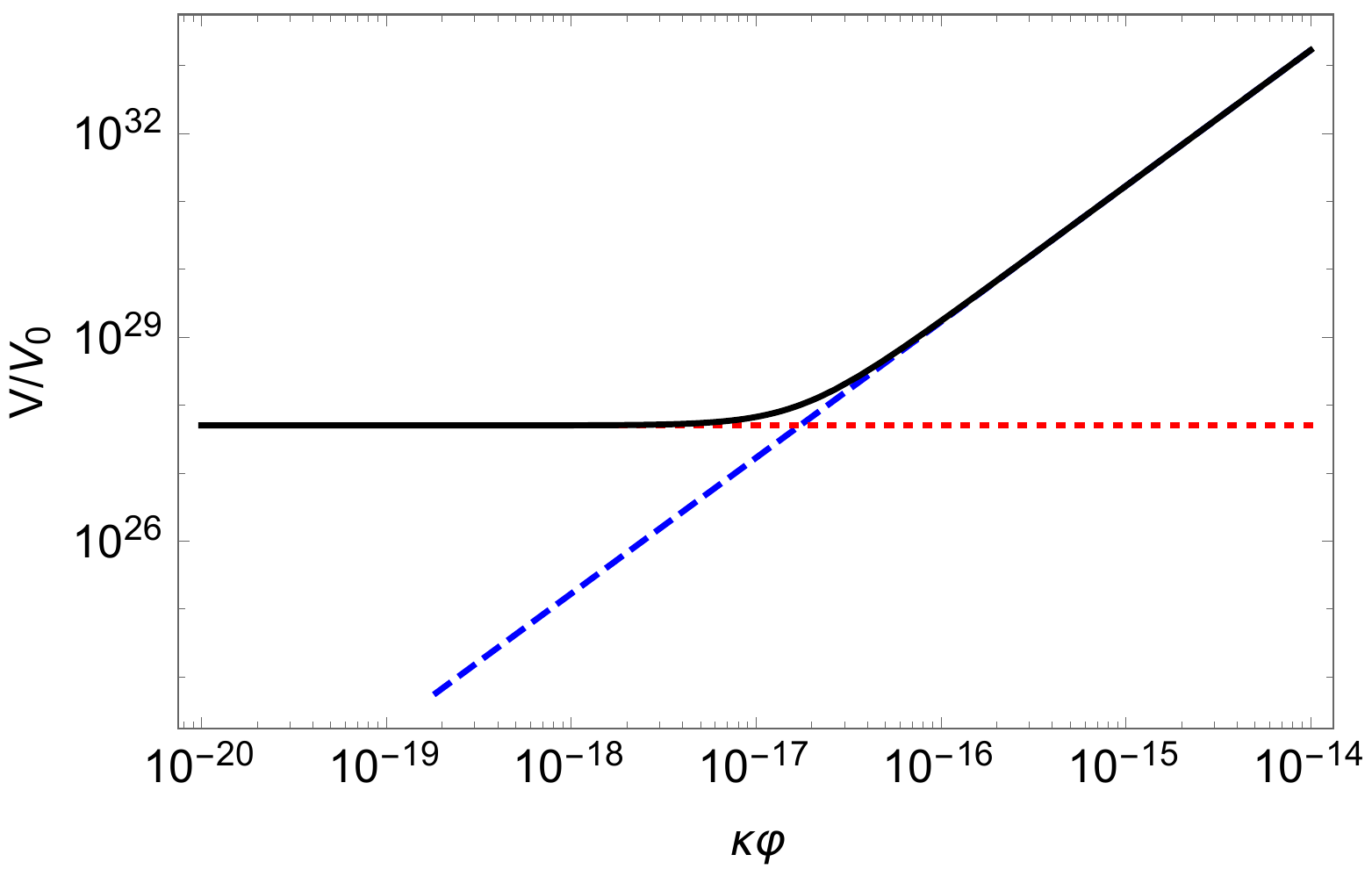}
\caption{The approximated effective potential in the BBN epoch (black solid line), which is the sum of the original potential (blue dashed line) and the matter contribution (red dotted line), with $-T^\mu_\mu(\mathrm{BBN})/\rho_\Lambda\sim10^{29}$.}
\label{Fig: Effective_potential_BBN_minimum}
\end{figure}

Secondly, we compare the approximated potential in Eq.~\eqref{Simplified potential} with the original potential in Eq.~\eqref{Original potential} 
(without matter contribution). See Fig.~\ref{Fig: Potential_original_approximated}. 
We find that those two potential forms fit well in the positive region, while there is a large gap in the negative region. 
So we should set the approximated potential to be zero in the negative region. 
An easy way is to multiply it with Heaviside step function $\Theta(\varphi)$,
\begin{align}
V(\varphi)
&\approx \frac{1}{12\kappa^{2}}\frac{(\kappa\varphi)^{2}}{\alpha\left(1+2\sqrt{1/6}\kappa\varphi\right)^{2}}\Theta(\varphi)
\,, \\
V_{,\varphi}(\varphi)
&\approx 
\frac{1}{6\kappa\alpha}\frac{\kappa\varphi}{\left(1+2\sqrt{1/6}\kappa\varphi\right)^{2}}\Theta(\varphi) \,,
\label{heaviside-type}
\end{align}
where we have Taylor expanded the exponential part for later convenience.
We can find that the minimum locates at around 
$\kappa \varphi \sim 10^{-17}$ in the BBN epoch as shown in Fig.~\ref{Fig: Effective_potential_BBN_minimum}. 
Note that the mass formula \eqref{Scalaron mass-analytic} holds
because the potential minimum always locates in the region $\varphi>0$.

Now that we have confirmed reliably approximated analytical expressions for the 
effective potential (Eqs.~\eqref{Simplified potential} and \eqref{heaviside-type}),  
we shall substitute them  
into Eq.~\eqref{Eq: Scalaron_evolution} to study the scalaron evolution. 
By assuming the radiation dominance, $H=\frac{1}{2t}$, 
the equation of motion Eq.~\eqref{Eq: Equation of motion} thus goes like
\begin{align}
& \tau^{2}\ddot{\phi}+\frac{3}{2}\tau\dot{\phi}+\frac{\tau^{2}}{6}\frac{\phi}{\left(1+2\sqrt{1/6}\phi\right)^{2}}\Theta(\phi) 
\notag\\
& 
-\frac{\xi}{\sqrt{6}}\left(1-2\sqrt{1/6}\phi\right)\left(\frac{3}{4}-\frac{1}{2}\tau^{2}\dot{\phi}^{2}-\tau^{2}\frac{\phi^{2}}{12\left(1+2\sqrt{1/6}\phi\right)^{2}}\Theta(\phi)\right)=0\,,
\label{Scalaron evolution}
\end{align}
 where $\phi=\kappa\varphi$ and $\tau=t/\sqrt{\alpha}$, both of which are dimensionless.

\begin{figure}[!htbp]
\centering
\includegraphics[scale=0.7]{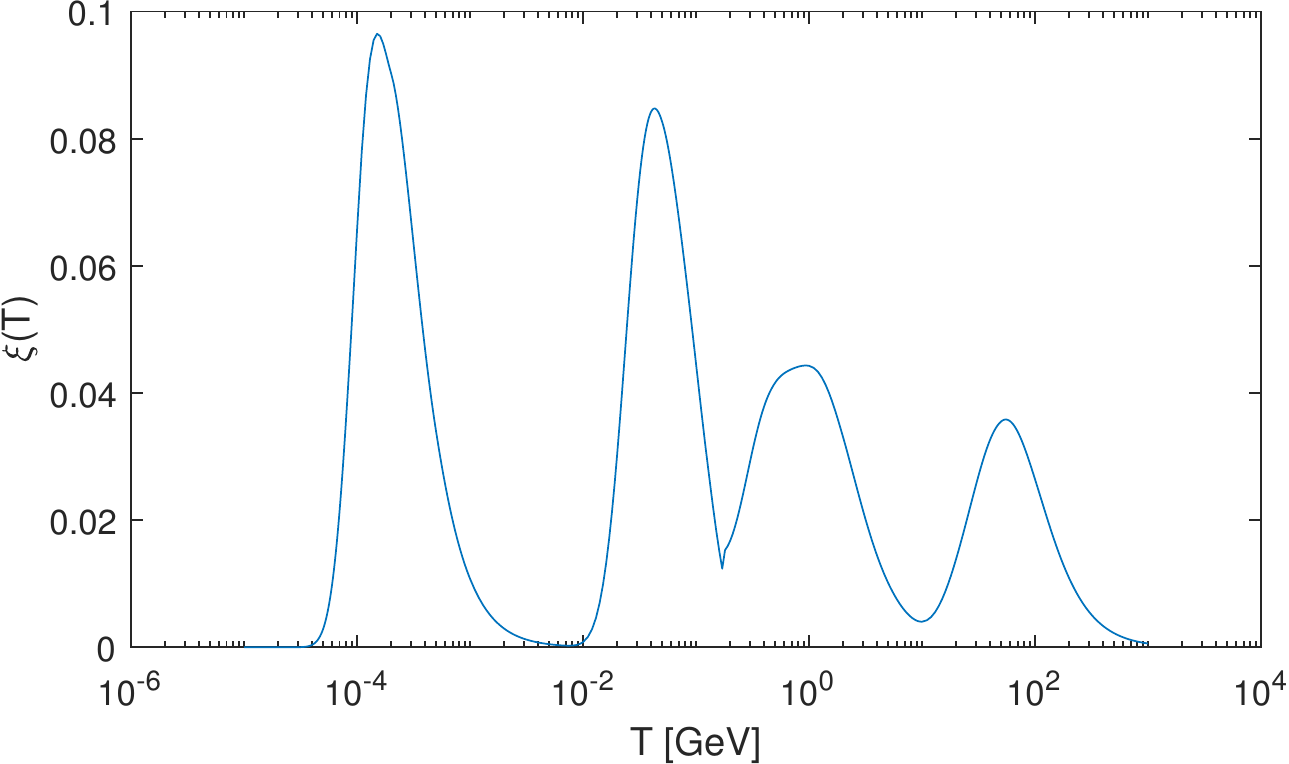}
\caption{The plot of $\xi(T)=-T^\mu_\mu/\rho$ as a function of temperature $T$.}
\label{Fig: xi}
\end{figure}

Because of the successive decouplings of SM species 
during the 
time evolution in the early universe (scaling $T\sim 100$[GeV] down to $\sim 1$[MeV]), 
$\xi(T)$ is expected to roughly oscillate between $0.01$ and $0.1$~\cite{Erickcek:2013oma, Katsuragawa:2017wge, Belokon:2018hrn}, as depicted in Fig.~\ref{Fig: xi}
which has been drawn by using the same evaluation of the $\xi(T)$ as in 
the literature~\cite{Katsuragawa:2017wge}.
Though the $\xi$ depends on the time in a nontrivial way, 
to get a rough insight,  
we shall take it to be constant in time. 
According to the form of the time evolution Eq.~\eqref{Scalaron evolution} with 
the vicinity of the potential minimum $(\phi \sim 0)$ focused on, 
this brute ansatz could be operative 
as long as the time scale is large enough compared to 
the contribution from 
the $\alpha R^2$ term, i.e. $\tau \gtrsim 1$. 

\begin{figure}[!htbp]
\centering
\includegraphics[scale=0.7]{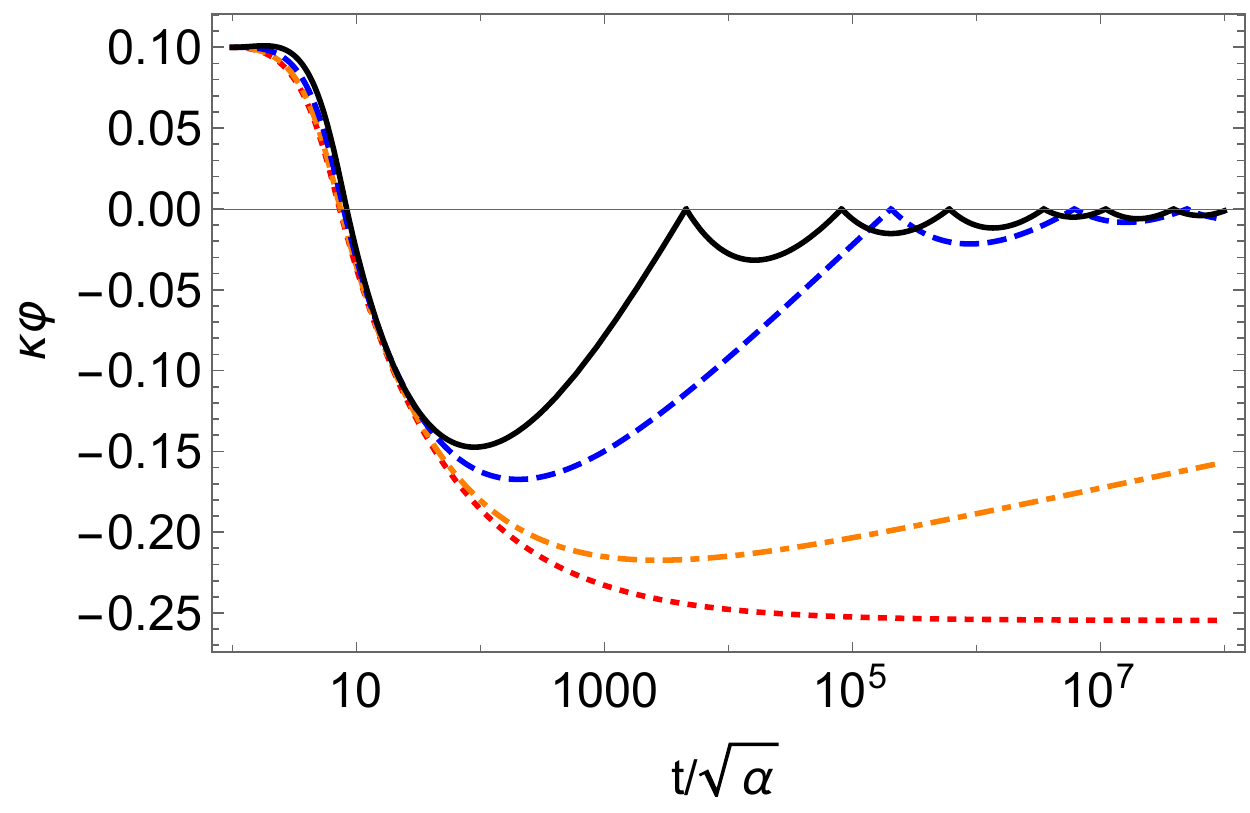}
\caption{The evolution of scalaron $(\phi=\kappa \varphi)$ in the normalized time $\tau=t/\sqrt{\alpha}$ 
for $\xi=0.1$ (solid line), $\xi=0.05$ (blue dashed line), $\xi=0.01$ (orange dot-dashed line), and $\xi=0$ (red dotted line), 
with the boundary condition $\phi(\tau=1)=0.01$ and $\dot{\phi}(\tau=1)=0$. }
\label{Fig: Scalaron_time}
\end{figure}

Figure~\ref{Fig: Scalaron_time} shows the time evolution of the scalaron 
with the initial condition $\phi(\tau=1) = 0.1$ and $\dot{\phi}(\tau=1)=0$, 
for $\xi=0.1, 0.05, 0.01$, in comparison with the case without matter contribution ($\xi=0$). 
Thus, with the matter contribution, 
the scalaron experiences a prompt falling and a slow recoil in the initial time evolution, and then it bounces at around zero to oscillate with damping for a larger time scale.  
Of importance is to notice that 
this is the charactereistic feature {\it universally} realized 
in a class of $F(R)$ gravity, as listed in Eqs.~\eqref{HS} - \eqref{T}.

This behavior can be understood by  
looking back Fig~\ref{Fig: Effective_potential_BBN}:  
the scalaron falls from a sharp and high potential barrier when it moves from
 the positive region, then 
turns to the negative region, 
where the matter contribution becomes dominant. 
Thus the scalaron decelerates till stop due to the matter contribution $T^\mu_\mu$ and Hubble friction $H(t)$. 
Then it will go back and will be blocked by the potential barrier, 
which is built very close to the origin, 
that is how it begins to bounce from the potential minimum.

The eminent turnaround takes place near $\tau \sim 10^2$ (i.e. $t\sim 10^{-11}$\ [s]
with $\alpha = 10^{22}\ [\mathrm{GeV^{-2}}]$), 
which correspond to a far past time even compared to the (expected) 
electroweak phase transition epoch around $\tau \sim 10^5$  
$(t\sim 10^{-8}$\ [s]), where the matter contribution can be highly 
model-dependent in a sense of possible models beyond the SM of 
particle physics.  
In the present study, we have simply assumed the order of $\xi $ does not alter, and the $\xi$ is not going to be zero.  
Note that the short periodicity and small amplitude of the oscillating scalaron field around the potential minimum at $\kappa \varphi_{\min} \sim 0$ allows Eq.~\eqref{Eq: Friedmann equation} to be almost reduced to the ordinary Friedman equation.
The above result is consistent with our assumption $H=\frac{1}{2t}$ in solving the equation of motion Eq.~\eqref{Scalaron evolution}.
As we will see in the next section,
the scalaron field affects observables in the early universe through the Hubble parameter.
Hereafter, we shall assume the radiation-dominant Hubble parameter to be evaluated by the relativistic species of the SM particles.

We note that Figure~\ref{Fig: Scalaron_time} also shows that the difference in the Jordan and Einstein frames is almost negligible.
Because $\mathrm{e}^{\sqrt{1/6} \kappa \varphi} \sim 1$,
the Weyl transformation in Eq.~\eqref{eq: Weyl transformation} implies $\tilde{g}_{\mu\nu} \sim g_{\mu\nu}$ during the BBN epoch.
Furthermore, the factor $\mathrm{e}^{3 \sqrt{1/6} \kappa \varphi}$ in the field redefinition or non-conservation of the matter fields do not significantly affect our results
although we have used the matter fields defined in the Jordan frame so far.
We will discuss the frame-difference in detail later.

\section{BBN versus scalaron}
\label{sec4}

Specifying $F(R)$ gravity in the early universe,
we have formulated the simplified model of $F(R)$ function and the trace of the energy-momentum tensor
based on the observation that the effective potential of the scalaron is governed by the $R^{2}$ term and matter contributions through the chameleon mechanism.
The numerical analysis of the scalaron dynamics in the FLRW space-time 
shows the damping oscillation of the scalaron in the early universe.
In the following analysis, 
from the oscillating behavior in the earlier universe,
we deduce the scalaron dynamics in the successive BBN epoch. 
Considering the scalaron couplings to the BBN and the frame-difference arising from the Weyl transformation,
we examine the BBN constraints on the scalaron.

First of all, 
it should be noted that 
the scalaron has to be non-relativistic in the BBN epoch, otherwise 
the successful nucleation processes would be spoiled 
(for a review on light particle contributions to the BBN, 
see the review part in Particle Data Group~\cite{Tanabashi:2018oca}). 
It gives a universal bound for the modified gravity scenario to survive,  
including the class of models listed in Eqs.~\eqref{HS} -~\eqref{E}. 
As has been demonstrated in~\cite{Katsuragawa:2017wge}, 
the scalaron mass evolves as an almost constant in time (temperature) around the BBN
where our approximated model Eq.~\eqref{R^2_approx} works.
In fact, we can easily check it from 
Eq.~\eqref{Scalaron mass-analytic} when $\alpha$ is small enough to satisfy the fifth-force constraints: 
\begin{align} 
m(\varphi_{\min}) \simeq \sqrt{\frac{1}{6 \alpha}} 
\,.  
\end{align}
Since the BBN took place at around $T \sim 1~[\mathrm{MeV}]$, 
for the scalaron to be non-relativistic, the parameter $\alpha$ thus gets constrained to be
\begin{align} 
 \alpha \lesssim 2 \times 10^5~[{\rm GeV}^{-2}]
\,, \label{alphacons}
\end{align}
which is more stringent than the bound that the fifth force has currently placed.

\subsection{Chameleon dark matter in BBN epoch}

A typical time scale for the BBN ($t \sim 10^{2}$[s]) corresponds to $\tau \sim 10^{15}$ in our setup. 
Thus, we can study the scalaron dynamics, which connects to the results in the previous section,  
by taking the large $\tau$ limit (with $|\phi| \ll 1$) in Eq.~\eqref{Scalaron evolution}.
In such a limit, one can approximate Eq.~\eqref{Scalaron evolution} as follows: 
\begin{align} 
\tau^{2} \ddot{\phi} + \frac{3}{2} \tau \dot{\phi} + \frac{\tau^{2}}{6} \phi \Theta (\phi) - \frac{3 \xi}{4\sqrt{6}} \approx 0\,,
\label{Approx}
\end{align}
for the small scalaron velocity $|\dot{\phi}| \ll 1$.
Eq.~\eqref{Approx} implies that the matter contribution gives a perturbative effect on the harmonic oscillation of the scalaron, 
and the Hubble friction term provides the damping factor (hence it is bounced), as long as the $\xi$ is small enough (like $\xi \sim 0.01 - 0.1$), 
but nonzero. 
Thus, we can deduce that in the BBN epoch, 
the scalaron is present as a stable field at the vacuum close to the origin of the potential ($\phi = \kappa \varphi \sim {\cal O}(10^{-17})$)
and keeps undergoing a slow damping oscillation, as seen in the case with orange dot-dashed line in Fig.~\ref{Fig: Scalaron_time}.

One can analytically solve Eq.~\eqref{Approx} in $\phi>0$ and $\phi<0$ regions individually.
For $\phi>0$, $\Theta(\phi)=1$, and thus 
the general solution of the homogeneous part is given by a linear combination of the Bessel functions of the first and second kinds with the negative power 
damping,  
$\phi(\tau) = c_{1} \tau^{-1/4} J_{1/4}(\tau/\sqrt{6}) + c_{2} \tau^{-1/4} Y_{1/4} (\tau/\sqrt{6})$ where $c_1$ and $c_2$ are integral constants. 
A particular solution of the inhomogeneous equation also gives the damping behavior in terms of the generalized hypergeometric function 
${}_{1}F_{2}({1/4}, \{5/4, 5/4 \}, - \tau^{2}/24)$
and the Meijer G-function with the negative power of $\tau$.
Thus, the scalaron shows the damping oscillation naturally in the 
$\phi>0$ region. 
For $\phi<0$, $\Theta(\phi)=0$, and thus the third term in Eq.~\eqref{Approx} disappears.
The equation generates a solution $\phi = (\xi/2)(\sqrt{3/2}) \ln(\tau) - 2c_{3}/\sqrt{\tau} + c_{4}$, where $c_{3}$ and $c_{4}$ are integral constants.

\begin{figure}[!htbp]
\centering
\includegraphics[scale=0.7]{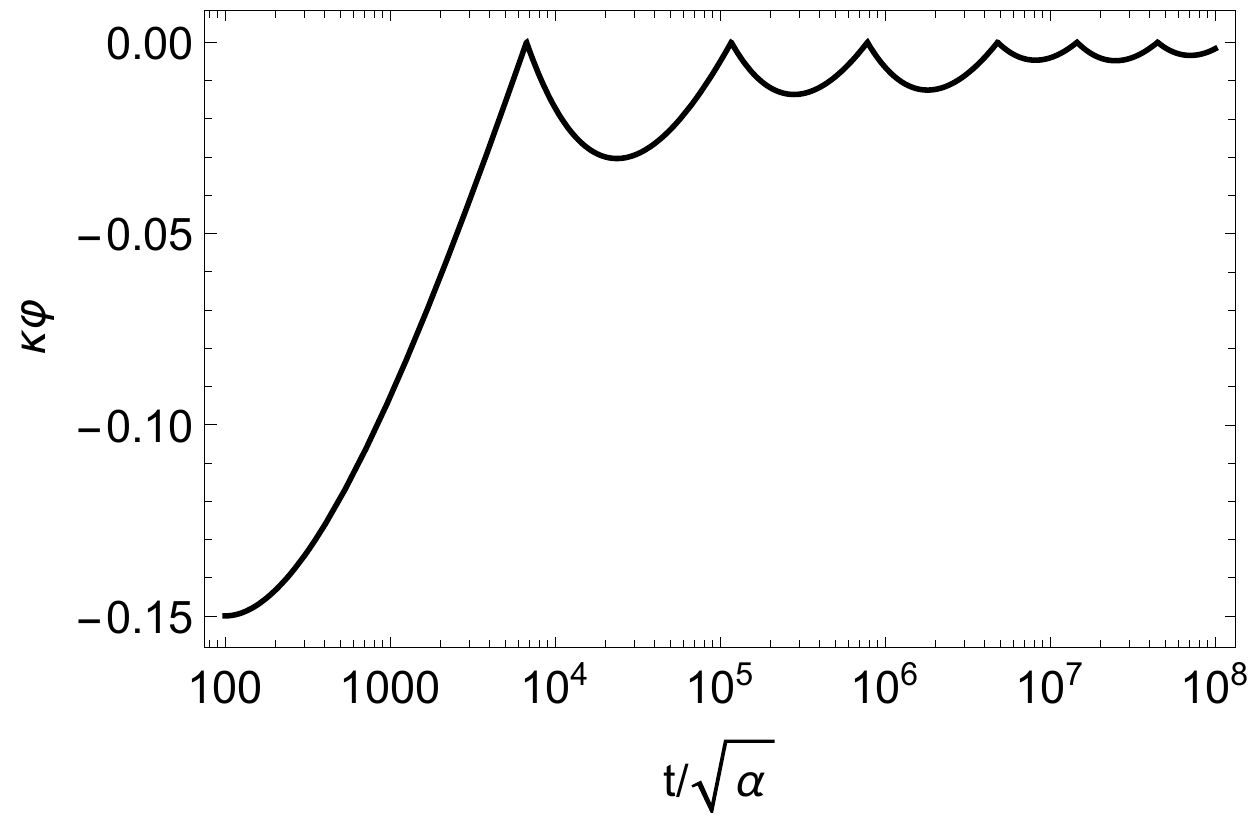}
\caption{
The evolution of scalaron in the large $\tau$ limit for $\xi=0.1$ corresponding 
to the solution of Eq.~\eqref{Approx}. 
The boundary condition $\phi(\tau=10^{2}) = -0.15$ and $\dot{\phi}(\tau=10^{2})=0$ are chosen from Fig.~\ref{Fig: Scalaron_time}. 
}
\label{Fig: Approx_Scalaron_time}
\end{figure}

Combination of the above analytic solutions allows us to understand the scalaron damping oscillation in the following way: 
When the scalaron rolls down in the potential in the $\phi>0$ region and reaches at $\phi=0$,
$\dot{\phi}$ will get a negative value to enter the $\phi<0$ region. 
A numerical analysis with the boundary condition $\phi<0$ at $\phi=0$
shows that the scalaron deceleratedly climbs the potential and stops.
When the scalaron rolls down in the potential in the $\phi<0$ region and comes back to $\phi=0$,
$\dot\phi$ will get a positive value to re-enter $\phi>0$ region.
Thanks to the damping oscillation in  the $\phi>0$ region,
the scalaron will go back to $\phi=0$ again with a negative velocity after an oscillation with very tiny (unseeable in our figures) amplitude.
The scalaron repeats the above and shows the damping oscillation in total, which is in perfect consistent with Fig.~\ref{Fig: Scalaron_time}.
We also plot a numerical result for Eq.~\eqref{Approx} in Fig.~\ref{Fig: Approx_Scalaron_time}, which shows the precise agreement with 
the full numerical analysis in Fig.~\ref{Fig: Scalaron_time}.

Note that the amplitude of the scalaron oscillation depends on the initial condition of the scalaron field. 
We have assumed that the scalaron is staying close to the potential minimum in the cosmic history,
and that the $R^2$ inflation scenario with a large $\varphi$ field is separated from our present interest.
The above is a natural assumption and allows the scalaron to roll down the potential in the earlier universe
if we incorporate the all cosmic history of the scalaron field. 
Moreover,
if the amplitude of the oscillation is small enough in the earlier universe,  
as shown in Fig.~\ref{Fig: Scalaron_time}, 
it is natural to expect following Eq.~\eqref{Approx} 
that the amplitude of the oscillation should be smaller than the order of unity in the BBN epoch. 

A similar argument was done in Ref.~\cite{Brax:2004qh}.
In the case that the initial value of $\phi$, $\phi_{i}$,  
locates in the region $\phi<0$,
the potential term induced by the $R^{2}$ term is negligible, 
and the matter contribution $T^{\mu}_{\mu}$ and the Hubble friction $3H\dot{\varphi}$ dominate in the equation of motion,
which forces the scalaron to get closer to the potential minimum.
On the other hand,
in the case that the $\phi_{i}$ locates in the region $\phi>0$,
the original potential term including the $R^2$ term dominates, and it makes the scalaron rolled down to the potential minimum.
Those two effects may justify our assumption that the scalaron locates around the potential minimum in the early universe,
which also does not allow us to set the unnaturally large initial value of the scalaron.

Keeping such an oscillation behavior of the scalaron in our mind,
we shall next discuss the frame-difference between the Jordan and Einstein frame. 
Because the matter sector is introduced in the Jordan frame, all dimension-full observables receive the effect of the Weyl transformation. 
According to the oscillating solution of the scalaron as in Eq.~\eqref{Approx}, 
we can naturally expand the scalaron field as
\begin{align}
\varphi = \varphi_{\min} + \delta \varphi\,, 
\end{align}
where $\varphi_{\min}$ and $\delta \varphi$ denote the field value at the potential minimum and the fluctuation of the scalaron field. Then the exponential form of the scalaron field in the Weyl transformation can be rewritten as
\begin{align}
e^{C \kappa \varphi} = e^{C \kappa \varphi_{\min}} e^{C \kappa \delta \varphi}\,, 
\end{align}
where $C$ is an arbitrary coefficient.

Since the fluctuation of the scalaron field is small enough, i.e., $\kappa \delta \varphi \ll 1$,
it can be expanded as follows:
\begin{align}
e^{C \kappa \varphi} \approx e^{C \kappa \varphi_{\min}} \left[ 1 + C \kappa \delta \varphi + \mathcal{O} (\kappa^{2} \delta \varphi^{2}) \right]\,.
\end{align}
The above equation allows having the perturbative picture for the scalaron couplings to the matter fields,
and the overall factor $e^{C \kappa \varphi_{\min}}$ is corresponding to the frame-difference associated with the Weyl transformation of the metric. 
We now note that the frame-difference is negligible because $\kappa \varphi_{\min}$ is almost equal to zero 
as read off from Fig.~\ref{Fig: Approx_Scalaron_time}. 
Consequently, there is almost no frame-difference in the observables such as the $S$-matrix elements including the mass spectra.
Hereafter, we will compute observables and constraints in the Einstein frame, which can be read off as those in the Jordan frame (see, \cite{Katsuragawa:2016yir}).

\subsection{Scalaron couplings to BBN}

Now we are ready to discuss the scalaron couplings to the BBN. 
The BBN constraints mainly arise from the precise measurement on the helium abundance $(Y_{{}^4{\rm He}})$. 
When there exists 
a nontrivial field configuration not involved in the standard BBN model, 
the $Y_{{}^4{\rm He}}$ can generically 
get modified through the corrections to the nucleon (neutron-proton) 
mass difference $\Delta m_N \equiv m_n - m_p$ and the neutron lifetime 
$\tau_n \sim [G_F^2 m_e^5]^{-1}$, with the electron mass $m_e$ and 
the Fermi constant $G_F$ 
(and possibly also to the effective degrees of freedom for relativistic 
particles at that time) because these are related to models of particle 
physics on which one works.   
The issue is, however, a bit complicated and somewhat nontrivial 
for the scalaron couplings to the nucleon, 
because 
the nucleon is a composite particle of quarks as a consequence of quantum chromodynamics, QCD.
For preparation to study the interactions between the scalaron and such a composite particle, 
we begin by reviewing the scalaron couplings to the elementary fermions, such as quarks and leptons, following Ref.~\cite{Katsuragawa:2016yir}.

Let us first consider the following Lagrangian of the massive fermion field $\psi$:
\begin{align}
\mathcal{L}_{F} \left( \gamma^{\mu}, \psi \right)
= 
i \bar{\psi} (x) \gamma^{\mu} \nabla_{\mu} \psi (x) - m_{F} \bar{\psi} \psi 
\, ,
\end{align}
where $m_{F}$ denotes the fermion mass,
and the generalized Dirac gamma matrix in the curved space-time satisfies $\left\{ \gamma^{\mu}, \, \gamma^{\nu} \right\} = - 2 g^{\mu \nu}$.
After the Weyl transformation of the metric $g_{\mu \nu} = e^{-2 \sqrt{1/6} \kappa \varphi} \tilde{g}_{\mu \nu}$,
the action of the fermion field is transformed as
\begin{align}
S
=&
\int d^{4}x \sqrt{-\tilde{g}} \, \mathrm{e}^{-4\sqrt{1/6}\kappa \varphi} 
\mathcal{L}_{F} \left( \gamma^{\mu}, \psi \right)
\nonumber \\
=&
\int d^{4}x \sqrt{-\tilde{g}} 
\left[ 
\mathrm{e}^{- 3\sqrt{1/6}\kappa \varphi} i \bar{\psi} \tilde{\gamma}^{\mu} \tilde{\nabla}_{\mu} \psi
- \frac{3 i }{2} \sqrt{\frac{1}{6}}\kappa \mathrm{e}^{ -3\sqrt{1/6}\kappa \varphi} \left( \partial_{\mu} \varphi \right) \bar{\psi} \tilde{\gamma}^{\mu} \psi  
\right. \nonumber \\
& \left. \qquad \qquad \qquad \qquad \qquad \qquad \qquad \qquad \qquad  \qquad \qquad
- \mathrm{e}^{-4\sqrt{1/6}\kappa \varphi}  m_{F} \bar{\psi} \psi 
\right] 
\label{FermionWeyl}
\, ,
\end{align}
where the Dirac gamma matrix in the Einstein frame is transformed as $\tilde{\gamma}^{\mu} = e^{- \sqrt{1/6}\kappa \varphi} \gamma^{\mu}$ for the consistency with the Weyl transformation of the metric.

Here, we can redefine the field $\psi$ to realize the canonical kinetic term:
\begin{align}
\psi \rightarrow \psi^{\prime} = \mathrm{e}^{-3/2 \sqrt{1/6} \kappa \varphi} \psi
\, .
\end{align}
For the above field redefinition, the action in Eq.~\eqref{FermionWeyl} is written in terms of $\psi^{\prime}$:
\begin{align}
S
=&
\int d^{4}x \sqrt{-\tilde{g}} 
\left[ 
 i \bar{\psi^{\prime}} \tilde{\gamma}^{\mu} \tilde{\nabla}_{\mu} \psi^{\prime}
- m_{F} \mathrm{e}^{-\sqrt{1/6}\kappa \varphi} \bar{\psi^{\prime}} \psi^{\prime}
\right] 
\, .\label{Dirac-Scalaron}
\end{align}
This reads the minimal coupling form between the scalaron and 
massive Dirac fermion fields. 

In a similar way, we may derive the scalaron coupling to a four-fermion interaction  
(with the coupling constant $G_{4f}$) 
\begin{align}
S_{4f}
=&
\int d^{4}x \sqrt{-\tilde{g}} 
\left[ 
G_{4f} \mathrm{e}^{+ 2 \sqrt{1/6}\kappa \varphi} (\bar{\psi^{\prime}} \psi^{\prime})^2 
\right] \,,
\label{4-fermi}
\end{align} 
which would be relevant to the neutron decay process. 

Thus, 
the interaction terms between the scalaron and massive (elementary) fermions arise like dilatonic scalar couplings reflecting the scale dimension for the corresponding interaction operators.  
Note also that
the scalaron coupling does not distinguish any charges and flavor of the leptons and quarks,
which gives us a naive expectation that 
the similar coupling forms show up for composite particles, such as the baryons and mesons after the Weyl transformation.

However, this is not the case for the nucleon: 
since the nucleon dynamics is highly nonperturbative in the sense of conventional perturbation theories, 
higher-dimensional nucleon interactions cannot simply be neglected. 
Among those non-minimal interaction terms, for instance, four-nucleon interaction terms of Lorentz vector(or scalar) type might be induced in QCD, 
which can arise along with the isospin breaking, 
such as $(\bar{N} \gamma_\mu \sigma^3 N)^2$ with $N$ and $\sigma^3$ being a nucleon doublet $(p,n)^T$ and the third component of the Pauli matrices, respectively. 
Generically, such interactions can generate nontrivial mass-split for the nucleon at quantum loop levels for the nucleon dynamics, and can also couple to the scalaron in a nontrivial way, 
due to the Weyl transformation and rescaling of the nucleon field, as demonstrated above. 
Thus, one naively suspects that the nonperturbative nucleon dynamics would drastically make the scalaron coupling to the nucleon mass modified from the minimal and the simplest form as in Eq.~\eqref{Dirac-Scalaron}. 
What is worse, 
one cannot naively suppress all higher-order terms beyond the four-fermion type, 
because of the nonperturbativity, 
so that the form of the scalaron coupling to the nucleon mass would be out of control at the quantum level. 
Note that this complexity is characteristic to the generic argument on the 
scalaron coupling to the nucleon, which does not come in the case of 
axion-like particle not associated with the Weyl transformation.

A way out of this complex issue can be hinted by noticing that the nucleon mass difference signals the explicit breaking of the chiral (isospin) symmetry for the nucleon. 
Since the observed size of the nucleon mass difference 
 is quite small compared to its mass,  
$\Delta m_N/m_N = (m_n - m_p)/m_{n/p} = {\cal O}(10^{-3})$, 
one can expand the $S$-matrix elements for nucleon scattering amplitudes including the mass shift in terms of the size of the isospin breaking. 
It is possible to demonstrate the above expansion based on the framework of the so-called chiral perturbation theory, where the multi-body nucleon interactions including the isospin breaking can be treated as perturbations compared to the leading-order correction set by the tree-level contribution from the quark mass difference to the nucleon mass.

As explicitly given in Appendix~\ref{appendix1}, 
the scalaron coupling to the nucleon mass difference $\Delta m_N$ can thus be evaluated based on the baryon chiral perturbation theory (BChPT). 
This BChPT is formulated in the Einstein frame but can be converted to the theory in the Jordan frame by taking into account a negligibly small discriminant exponential factor, as argued in the above. 
It will be ignorable compared to other uncertainties, e.g., in the BChPT.    
It turns out that at the nontrivial leading order (LO) of the perturbation, 
the scalaron affects the mass difference just like in a dilatonic coupling form, 
which coincides with the one read off from the minimal Dirac fermion-scalaron action  
in Eq.~\eqref{Dirac-Scalaron}: 
\begin{align}
\Delta m_N(\varphi)|_{\rm LO} 
&=
[m_n(\varphi)-m_p(\varphi) ]|_{\rm LO}
=(m_n-m_p) e^{-\kappa\varphi/\sqrt{6}}\,.
\notag \\ 
& 
\equiv 
\Delta m_N e^{-\kappa\varphi/\sqrt{6}}
\,. \label{LOmass}
\end{align}

Beyond the leading order corrections, the nucleon-mass splitting could get affected by interactions with pions and also by electromagnetic (EM) interactions with photon when the chiral symmetry is (in part) gauged. 
As in the literature~\cite{Gasser:1982ap}, 
among those next LO corrections, 
the EM corrections will be most significant in magnitude. 
Indeed, the total nucleon - proton mass difference is shared by 
the EM correction part with a ``pure QCD'' part 
(the up- and down-quark mass difference) like 
\begin{align} 
 \Delta m_N^{\rm total} 
 & \equiv \Delta m_N^{\rm EM} + \Delta m_N^{\rm QCD}  
  \notag \\ 
& \simeq \left( \{- 0.76\}_{\rm EM} + \{2.05\}_{\rm QCD}\right)\, [{\rm MeV}]  
\notag \\ 
& = 1.29 \, [{\rm MeV}]
\,, \label{GLformula}
\end{align}
where the pure QCD part has been extracted just by 
subtracting the observed value $\simeq 1.29$~[MeV]~\cite{Tanabashi:2018oca} by 
the EM term estimated assuming a one-photon exchange contribution 
based on the current algebra technique~\cite{Gasser:1982ap}. 
Eq.~\eqref{GLformula} implies that the EM correction pulls the mass difference down 
by about 37\%, so it is indeed somewhat a significant destructive interference, 
which implies that the scalaron coupling to the EM correction part needs to be 
taken into account as well.

In Eq.~\eqref{GLformula}, 
the $\Delta m_N^{\rm QCD}$ is identified as the $\Delta m_N$ in Eq.~\eqref{LOmass}. 
Regarding the EM correction, the scalaron can actually modify it through the EM scale anomaly $\sim e^2/(4 \pi)^2 
F_{\mu\nu}F^{\mu\nu}$, as discussed in~\cite{Katsuragawa:2016yir}. 
The scalaron thus makes the EM coupling $e$ modified as 
$e^2 \to e^2(\varphi) = e^{-\sqrt{1/6} \kappa \varphi} \cdot e^2$~\cite{Katsuragawa:2016yir} 
along with the beta function coefficient arising from the charged fermion loops. 
Note that the scaling factor for the EM correction part is the same as that for the quark mass difference ($\Delta m_N$) part. 
Thus, the scalaron just gives an overall scaling for the neutron - proton mass difference 
in total, in such a way that 
\begin{align} 
\Delta m_N^{\rm total}(\varphi) = \Delta m_N^{\rm total}  e^{-\sqrt{1/6} \kappa \varphi}
\,. \label{total:mN:varphi}
\end{align}  
We shall later on use this Eq.~\eqref{total:mN:varphi} to 
discuss a rough size of effects on the BBN.

As to the nucleon lifetime $\tau_n \sim [G_F^2 m_e^5]^{-1}$ 
which is saturated by the beta decay $n \to p + e + \bar{\nu}$, 
it is not due to the strong QCD, but the weak interaction process. 
Hence we do not need to apply the BChPT but scale it by the scalaron field factor of $e^{-\kappa \varphi/\sqrt{6}}$ with respect to the scale dimension of minus one, 
following the four-fermion interaction and elementary fermion mass terms in Eqs.~\eqref{Dirac-Scalaron} and~\eqref{4-fermi}. 
Thus we have 
\begin{align}
\tau_n(\varphi) = \tau_n \cdot e^{+ \kappa \varphi/\sqrt{6}} 
\,. \label{n-lifetime}
\end{align}

Now that we have derived the relevant formulae (Eqs.~\eqref{total:mN:varphi} and \eqref{n-lifetime})
for the scalaron effect on the 
BBN, in the next subsection we shall discuss the BBN constraint on the 
scalaron and the size of its sensitivity -- possibility to hunt the scalaron -- 
in the near future.

\subsection{BBN constraints}

We begin by looking at the generic formula for 
the helium abundance at the BBN epoch $Y_{{}^4{\rm He}}(t_{\rm BBN})$~\cite{Sarkar:1995dd}:  
\begin{align} 
Y_{{}^4{\rm He}}(t_{\rm BBN}) = 
Y_{{}^4{\rm He}}(t \to 0) \cdot e^{- t_{\rm BBN}/\tau_n} 
\,, \label{YHe}
\end{align}
in which for later concrete estimates, 
we will take $t_{\rm BBN} \simeq 180$[s], 
corresponding to $\simeq 1$~[MeV].   
The overall yield $Y_{{}^4{\rm He}}(t \to 0)$ is given by integrating 
the balance equation involving a couple of  $n \to p$ conversion processes 
($n \nu_e \to p e^-, ne^+ \to p \bar{\nu}_e, n \to p e^- \bar{\nu}_e$)~\cite{Sarkar:1995dd}: 
\begin{align} 
 Y_{{}^4{\rm He}}(t \to 0)
 = \int_0^\infty dy \frac{e^{y - K(y)}}{(1 + e^y)^2} 
 \,, 
\end{align}   
where $y=\Delta m_N/T$, and 
\begin{align} 
K(y) &= b \cdot \left[
\left( \frac{4}{y^3} + \frac{3}{y^2} + \frac{1}{y} \right) 
+ \left( \frac{4}{y^3} + \frac{1}{y^2} \right) e^{-y}
\right] 
\,, \notag \\ 
b &= 253 \cdot \sqrt{\frac{45}{4 \pi^3 g_*(t_{\rm BBN})}} \frac{\sqrt{2} M_\mathrm{pl}}{\tau_n \Delta m_N^2} 
\,,   
\end{align}
where $g_*(t_{\rm BBN})$ denotes the effective degrees of freedom for relativistic 
particles in the BBN era. 
Here we have assumed that 
the scalaron does not significantly affect the Hubble parameter $H$ in Eq.~\eqref{Eq: Friedmann equation} in the BBN epoch, and the $H$ can be evaluated by the fluid picture for the relativistic particles at that time $t_{\rm BBN}$. 
The former assumption would be reasonable: 
in Fig.~\ref{Fig: Scalaron_time} with a small enough matter contribution 
($\xi \lesssim {\cal O}(10^{-2})$),  
we have observed that the approximated time evolution Eq.~\eqref{Approx} 
describes the damping oscillation very well for a large $\tau_{\rm BBN} 
= t_{\rm BBN}/\sqrt{\alpha} \sim 10^{24}$. 
It implies that we can numerically neglect the implicit scalaron dependence in the Hubble parameter $H$ as in Eq.~\eqref{Eq: Friedmann equation}, and replace the $H$ as the standard BBN's.

As to the effective degrees of freedom $g_*(t_{\rm BBN})$, 
we will take 
$g_*(t_{\rm BBN})\simeq 3.36$ that is a standard value  
after the $e^+ e^-$ annihilation~\cite{Sarkar:1995dd}, 
because the scalaron acts like a non-relativistic particle to make no 
contribution to the background Hubble parameter, including 
the above argument on its numerically negligible time evolution effect.

Though the scalaron may not significantly affect the background 
Hubble evolution as argued above, 
the scalaron can be hunted by the BBN: 
Since the scalaron fluctuation in the BBN epoch 
can be controlled to be small enough, due to the presence of the balance between the $R^2$ term and the chameleon mechanism, 
the size of the deviations from the SM predictions in Eqs.~\eqref{total:mN:varphi} 
and \eqref{n-lifetime} can be small as well. 
In that case, the helium abundance will get a small shift to be modified from the one in Eq.~\eqref{YHe} by the small amount. 
Then we can easily see the correction arises in a perturbative series 
like 
\begin{align} 
Y_{{}^4{\rm He}}(t_{\rm BBN})|_{\varphi} 
= 
Y_{{}^4{\rm He}}(t_{\rm BBN})|_{\rm SM} 
\left[1 - 2.27 \left(\frac{\kappa \varphi}{\sqrt{6}} \right) + \cdots \right]
\,, 
\end{align}
with the higher order terms neglected. (Here we have used $\tau_n \simeq 880$[s] from 
Particle Data Group~\cite{Tanabashi:2018oca}.) 
According to the current status on the standard BBN prediction estimated 
using the latest Planck 2018 data 
 for the baryon number density fraction, 
we may quote $Y_{{}^4{\rm He}}(t_{\rm BBN})|_{\rm SM} \simeq 0.247$ (at 
95\% C.L.)~\cite{Aghanim:2018eyx}. 
Taking the 1 sigma range for the observed helium abundance, 
$Y_{{}^4{\rm He}}(t_{\rm BBN})|_{\rm obs} 
= 0.243 ^{+ 0.023}_{-0.024} $ 
(including gravitational lensing and baryon acoustic oscillation data and taking 
the best-fit number for the neutrino species, $N_{\rm eff}=3.046$ at present)~\cite{Aghanim:2018eyx}, 
we thus get the BBN bound on the scalaron field,  
\begin{align} 
 -0.05 < (\kappa \varphi) < 0.03
\,. \label{BBNcons}
\end{align}

In the previous section, 
we showed that thanks to the $R^2$ term and the chameleon mechanism, 
the scalaron always stays around the minimum of the effective potential,  
with the size of the amplitude
$|\kappa \varphi| 
= {\cal O}(10^{-3}-10^{-2})$, 
in the BBN epoch 
(see Figs.~\ref{Fig: Effective_potential_BBN} and~\ref{Fig: Scalaron_time}). 
We again note that the amplitude of the scalaron depends on the initial condition.
Considering the damping behavior of the scalaron field 
(as seen in Fig.~\ref{Fig: Scalaron_time}), 
we see that the scalaron fluctuation can naturally be small enough to survive the BBN constraint in Eq.~\eqref{BBNcons}, 
unless we impose the unnatural initial condition for the scalaron field.
Note the current observational accuracy on the helium abundance is roughly on the order of the allowed size for the oscillation amplitude for the scalaron $|\kappa \varphi| \lesssim 10\%$. 
Therefore, future updated measurements on 
the light element abundance as well as the baryon number 
density fraction would exclude 
a class of scalaron scenarios with somewhat a larger initial value $(\sim 0.1 -1\%)$ 
as in Fig.~\ref{Fig: Scalaron_time}, 
or might probe the hint for the presence of the scalaron in BBN through a significant deviation from the standard BBN model.

Thus, the BBN can be a definite hunter for the scalaron universally arising from a viable class of the $F(R)$ gravity having the irrespective low-energy modifications for the late-time cosmic acceleration. 
It is notable that the BBN can constrain the parameter $\alpha$
although we have fixed it as the experimental bound in the present analysis.
If $\alpha$ gets smaller (possibly due to updating the fifth-force experiments), 
the smaller the period of the scalaron oscillation becomes, the sooner the amplitude becomes damped.
Since the value of the scalaron field at the BBN epoch depends on $\alpha$,
we can thus utilize the observation of the helium abundance to constrain the high-energy modification for the $F(R)$ gravity through the $\alpha R^{2}$.  

\section{Conclusion}
\label{sec5}

Dark energy and dark matter are two of the biggest mysteries in cosmology and particle physics. 
Although they are usually studied separately, it is still possible to postulate that they have the same origin if one tries to resolve the phenomenological coincidence problem. 
In recent years, there are some attempts to incorporate the dark matter into the framework of modified gravity besides dark energy. 
In this paper, we have studied the unification of dark sector in the framework of $F(R)$ gravity and investigate the evolution of chameleon dark matter in the BBN epoch. 

We have introduced the second-order correction $R^2$ to cure the singularity problems which show up in generic models of the $F(R)$ dark energy.
We have confirmed that in the early universe, the model dependence of modifications for dark energy disappears, 
which is reasonable because the infrared dark energy has little effect on higher energy physics. 
Thus we have demonstrated that 
in the early universe, the model in $F(R)$ gravity which corresponds to the dark sector could be approximated as $R+\alpha R^2$, 
to find that the scalaron universally evolves in a way with bouncing oscillation, 
irrespective of the low-energy modification for the late-time cosmic acceleration (Fig.~\ref{Fig: Scalaron_time}).
Such a universal feature shows up due to the $R^2$ term and the chameleon mechanism characteristic to the scalaron.

We then have explored the cosmological and phenomenological consequences for the scalaron to present in the BBN epoch. 
We first have found a generic bound on the scalaron mass in the BBN epoch, 
to feedback to the constraint on the coupling strength for the $R^2$ term as in Eq.~\eqref{alphacons}.  
This bound is universally applicable to the viable modified gravity and more stringent than the one placed by the fifth force experiments.  
Hence the scalaron naturally develops a small enough fluctuation in the BBN epoch. 
By formulating the scalaron couplings to the BBN based on 
the BChPT, it is shown that the scalaron can have a good enough sensitivity 
for the BBN, and can avoid the current BBN constraint placed by the latest 
Planck 2018 data as shown in Eq.~\eqref{BBNcons}. 
The size of the detection sensitivity depends on the initial condition for the scalaron fluctuation in the early universe.  
We, therefore, expect that the BBN data with more accurate measurements for light element abundances as well as the baryon number density fraction can exclude, or probe some parts of the scalaron scenario.

At last, we will make several comments for future works.
In our analysis, we have approximated $\xi(T)$ to be constant 
in time ($\xi = 0.01, 0.05, 0.1$ as in Fig.~\ref{Fig: Scalaron_time}), 
which has given sufficient information about the scalaron field in the BBN, 
as far as the typical order of magnitude is concerned.
Although the temperature dependence of $\xi(T)$ has computed in~\cite{Katsuragawa:2017wge} and also in Fig.~\ref{Fig: xi} of the present paper, 
we still have theoretical ambiguities for the earlier or later cosmic history;
that is, we can include the various effects from models beyond the SM at the higher energy scale, and the decouplings and light elements at lower energy scales.
Two of the authors have embedded the scalaron dynamics into the scale-invariant extension of the SM and shown the vanishing $T_\mu^\mu$ at the electroweak phase transition.
Thus, $T_\mu^\mu$ goes from zero to nonzero, which would be the first kick to make the scalaron close to the potential minimum.
In this case, the argument about the initial condition should be examined to relate the scalaron physics with the possible (classical) scale-invariance in the very early universe.

Moreover, towards the complete analysis of the scalaron physics in the BBN epoch,
we have to address a nontrivially interacting system which comprises the scalaron and the other matters, 
instead of merely substituting the cosmic history of the other matters as external fields into the scalaron potential term.
It is mandatory to reveal the scalaron couplings to the composite particles, that is, nucleons and light elements in the early universe,
in which the BChPT approach we have employed in this paper shall play crucial roles.
Notably, such research also gives us potential tools to study theoretical predictions for the DM direct search experiments in the current universe.
Various experiments are designed to detect the dark matter candidate through the scattering with nucleus or electron in the atom. 
The detector in such experiments consists of, for instance, liquid xenon or argon, and the scalaron collides with the atomic nucleus.
Thus, as a branch of this work, 
the study on the scalaron coupling to the composite particles has a potential impact for the future works to reveal the scalaron dynamics in both early and current universe.

Finally, we comment on a problem of the chameleon mechanism in the early Universe. 
The earlier works \cite{Erickcek:2013oma,Erickcek:2013dea} based on the exponential model of the scalar-tensor theory have suggested that
the energetic quantum fluctuation of the scalaron field is produced by the interaction with the classical background in the following way: 
the scalaron field acquires the large velocity, called the surfer solution, after the final matter kick, so that the scalaron climbs its sharp potential.
Because the scalaron mass rapidly changes on a short time scale,
the non-adiabatic variation of the scalaron field shows up, which causes the energetic excitation modes.
As a result, such quantum effects dominate the scalaron dynamics and results in the breakdown of the calculability prior to the BBN epoch.

We can qualitatively compare our current work with the above earlier works,
which may allow us to avoid the catastrophic scenario.
In our model, 
we have a steep potential similar to the exponential potential thanks to the $R^2$ correction (see Fig.~\ref{Fig: Effective_potential_BBN}),
which depends on the parameter $\alpha$.
When we take a small $\alpha$, 
the potential can be high enough to prevent the scalaron field from going far away from $\varphi = 0$, 
and thus, the scalaron field can stay around $\varphi=0$.
By using Eq.~\eqref{Second derivative of simplified potential},
one can find that the scalaron mass does not change rapidly in our model when $\varphi \sim 0$. 
Therefore, in contrast to the exponential model, 
our model can potentially avoid the non-adiabatic production of quantum fluctuations even though the scalaron is accelerated by the matter kick.

The above consideration could tell us the new bound on the parameter $\alpha$,
and we can demonstrate it by evaluating the surfer solution and the chameleon velocity of our model in a way similar to the earlier works.
In relation to the non-adiabatic process in the catastrophic scenario,
the preheating at the potential minimum can also nonperturbatively produce the quantum fluctuation of chameleon 
although such a self-producing is already known to be technically difficult to compute because of the nonlinearlity.
Furthermore, the reheating to create the standard model particles by the nonperturbative process would be interesting.

\begin{acknowledgments}
We are grateful to Masato Yamanaka for useful discussions related to the BBN bounds,
and to Seishi Enomoto for useful comments about the nonperturbative particle production.
T.K. is supported by International Postdoctoral Exchange Fellowship Program at Central China Normal University
and project funded by China Postdoctoral Science Foundation 2018M632895. 
S.M. is supported in part by the National Science Foundation of China (NSFC) under Grant No.~11747308, and the Seeds Funding of Jilin University.
S.M. thanks for the hospitality of Institute of Astrophysics, Central China Normal University, where the present work has been in part done. T.Q. is supported by the National Science Foundation of China (NSFC) under Grant No.~11875141 and No.~11653002.
\end{acknowledgments}

\appendix

\section{Baryon chiral perturbation at leading order}
\label{appendix1}

To formulate the mass difference coupled to the scalaron in the Einstien frame, in this Appendix, we introduce the techniques in the hadron physics, which allows one to incorporate isospin breaking effects in a systematic way, called the baryon chiral perturbation theory (BChPT).

\subsection{Chiral Lagrangian for meson sector}

We begin by formulating the ChPT for the lightest meson sector, 
where  the chiral $SU(2)_L \times SU(2)_R$ symmetry 
is encoded in the nonlinear transformation law of 
the pion fields $\pi= \pi^a \tau^a$ with the generator of $SU(2)$ group, $\tau^a = \sigma^a/2$ (normalized as ${\rm tr}[\tau^a \tau^b] = \delta^{ab}/2$)  
and the Pauli matrices $\sigma^a$ ($a=1,2,3$). 
The chiral field $U$ parameterizing the pion fields is then expressed  
by a product of representatives of the coset space $G/H=[SU(2)_L \times SU(2)_R]/SU(2)_V$, 
$\xi_L$ and $\xi_R$ (with $\xi_R^\dag = \xi_L$), as 
$
 U = \xi_L^\dag \cdot \xi_R = e^{2 i \pi/f_\pi}  
$, 
with the pion decay constant $f_\pi$. 
The representatives $\xi_{L/R}$ transform under the $G$-symmetry 
nonlinearly with respect to the pion fields, like 
$\xi_{L/R} \to h(\pi, g_L, g_R)\cdot \xi_{L/R} \cdot g_{L/R}^\dag$ 
($g_{L/R} \in G= SU(2)_{L/R}$ 
and $h(\pi, g_L. g_R) \in H=SU(2)_V$), 
and hence $U \to g_L \cdot U \cdot g_R^\dag$.

It is convenient to introduce 1-forms (called Maurer-Cartan 1-forms) in 
constructing the chiral Lagrangian, defined as 
$\alpha_\mu^{L/R} \equiv - i \partial_\mu \xi_{L/R} \cdot \xi_{L/R}^\dag$, 
and those chiral-projected ones: $\alpha_\mu^{\perp, ||} \equiv (\alpha_\mu^{R} \mp \alpha_\mu^L)/2$. Note that among these 1-forms, 
the $\alpha_\mu^\perp$ transforms homogeneously under the $G$-symmetry, 
$\alpha^\perp_\mu \to h(\pi, g_L, g_R) \cdot \alpha^\perp_\mu \cdot h^\dag(\pi, g_L, g_R)$, 
while others do like a ``gauge field'' (called chiral connection field), 
 $\alpha^{L,R,||}_\mu \to h(\pi, g_L, g_R) \cdot \alpha^{L,R,||}_\mu \cdot h^\dag(\pi, g_L, g_R) 
 + i h(\pi, g_L, g_R)  \partial_\mu h^\dag(\pi, g_L, g_R)$. 
Thus the chiral Lagrangian invariant under the $G$-symmetry 
can be written in terms of those 1-forms, and 
at the leading order of derivative terms (of ${\cal O}(p^2)$) we find 
\begin{align} 
{\cal L}_{(2)} = f_\pi^2 {\rm tr}[ \alpha_\mu^\perp \alpha^{\mu \perp} ]
\,,  \label{Lag:p2}
\end{align} 
 where at this moment 
 the kinetic terms of the pion fields have been canonically normalized.  

To incorporate the explicit-chiral symmetry breaking effect reflecting 
the current quark mass terms in the underlying QCD (arising from 
the electroweak symmetry breaking via the Higgs), 
one may introduce a spurion field $\chi$, which makes 
the current quark mass term chiral-invariant form, like $\bar{q}_L \chi q_R + \bar{q}_R \chi^\dag q_L$, by allowing the $\chi$ to transform in the same way as the chiral 
field $U$ does. 
The $\chi$ is then assumed to develop the ``vacuum expectation value'' 
$\langle \chi \rangle = 2 B_0 {\cal M}$ with ${\cal M}$ = diag$\{m_u, m_d\}$, so that the 
chiral symmetry is explicitly broken consistently with the underlying QCD. 
Here the overall coefficient $B_0$ will be related to the chiral condensate later. 
Then one may have the following term, corresponding to the explicit breaking by 
the current quark mass,  
 in a chiral invariant way: 
\begin{align} 
{\cal L}_\chi 
= \frac{f_\pi^2}{4} {\rm tr}[\chi^\dag U + U^\dag \chi] 
=\frac{f_\pi^2}{4} {\rm tr}[\hat{\chi}^\dag + \hat{\chi}] 
\,, \label{Lag:chi}
\end{align}
where in the second equality the $\hat{\chi}$ reads $\hat{\chi}=\xi_L \cdot \chi \cdot \xi_R^\dag$, which transforms like $\hat{\chi} \to h(\pi, g_L, g_R) \cdot \hat{\chi} \cdot h^\dag(\pi, g_L, g_R)$. Expanding the ${\cal L}_\chi$ in powers of the $\pi$ fields, one can readily 
find the pion mass, $m_\pi^2 = \frac{B_0}{2}(m_u + m_d)$, which can be compared to 
the Gell-Mann-Oakes-Renner relation based on the current algebra $(m_\pi^2 f_\pi^2 = - \langle \bar{q}q \rangle (m_u + m_d))$ to derive $B_0 = - 2 \langle \bar{q}q \rangle/f_\pi^2 $.  

The chiral perturbation theory is thus constructed from the 1-form $\alpha_\mu^\perp$ and the spurion field $\hat{\chi}$, by assuming the pion mass $m_\pi$ to be much smaller than a typical chiral breaking  scale $\Lambda_\chi \sim 4 \pi f_\pi$ (which can be estimated by equating tree-level and one-loop contributions for pion scattering amplitudes arising from the interaction terms in ${\cal L}_{(2)}$), and respecting the low-energy theorem for the chiral symmetry which leads to vanishing amplitudes involving pions when the transfer momentum $p$ ($\sim$ soft pion mass $m_\pi$) is set to zero. Namely, the theory is built based on the derivative expansion 
with the order counting rule as 
$\partial_\mu \sim m_\pi \sim {\cal O}(p)$, so that the 1-form $\alpha_\mu^{\perp} 
\sim {\cal O}(p)$ and the spurion field $\hat{\chi} \sim {\cal O}(p^2)$. 
Thus one realizes that 
the leading-order chiral Lagrangian of ${\cal O}(p^2)$
is given by ${\cal L}_{(2)}$ in Eq.~\eqref{Lag:p2}
and ${\cal L}_\chi$ in Eq.~\eqref{Lag:chi}
respecting the chiral symmetry, and also the charge $(C)$ and parity $(P)$ conjugations.
The $C$ and $P$ transformations can be set to the building blocks, 
following the corresponding quark current form coupled to the underlying     QCD, 
so that $\alpha_\mu^\perp \leftrightarrow - \alpha^{\mu \perp}$ and $\hat{\chi} 
\leftrightarrow \hat{\chi}^\dag$ under $P$, 
$\alpha_\mu^\perp \leftrightarrow (\alpha_\mu^{\perp})^T$ and $\hat{\chi} 
\leftrightarrow \hat{\chi}^T$ under $C$. 
The higher derivative terms can systematically be incorporated 
to renormalize loop corrections arising from the lower derivative interactions 
order by order~\cite{Gasser:1983yg, Gasser:1984gg}.

\subsection{Inclusion of nucleon in chiral perturbation theory} 

Nucleon is heavy enough to be comparable with the chiral breaking scale $\sim 4 \pi f_\pi 
\sim 1$ GeV, and does not have the soft mass limit (never goes to zero in the 
chiral limit) unlike the pion, 
so it involves the problematic issue to incorporate the nucleon into 
the chiral perturbation theory, which would naively lead to the loss of the systematic 
order counting as presented in the meson sector.  
Indeed, it has been founded so far that the higher loop corrections actually bring significant cancellations in the baryon mass spectra, although the estimated masses tend to be close to the result from lattice simulations. (For  
the BChPT up to the next-to-next-to-next-to leading order 
v.s. lattice simulations, e.g. see~\cite{Ren:2012aj}, and references therein).   
In the present study, we shall work on the nontrivial-leading order terms in the BChPT, and try to give a rough order of estimate on the effect from the chameleon in the nucleon physics.

We introduce the chiral-nucleon fields $\hat{N}_{L,R}$ dressed by the ``pion clouds" of  
the representatives $\xi_{L,R}$ as $\hat{N}_{L,R} = \xi_{L,R} \cdot N_{L,R}$, where 
the $N_{L,R}=(p, n)^T_{L,R}$ denote the undressed ones transforming under the $G$-symmetry 
like $N_{L,R} \to g_{L,R} N_{L,R}$ 
and hence $\hat{N}_{L,R} \to h(\pi, g_L, g_R) \hat{N}_{L,R}$.  
The chiral- (and also $C$- and $P$-) invariant nucleon Lagrangian based on the nonlinear realization 
is at the leading order of ${\cal O}(p)$
 written to be  
\begin{align} 
{\cal L}_{(1)}^N = \bar{\hat{N}} i \gamma^\mu (\partial_\mu - i \alpha_\mu^{||} ) \hat{N} 
+ g_A \bar{\hat{N}} i \gamma^\mu \gamma^5 \alpha_\mu^\perp \hat{N} 
- m_0 \bar{\hat{N}}\hat{N} 
+ \cdots 
\,,    \label{Lag:N}
\end{align}
where ellipses stand for higher order terms in the nucleon field (e.g. four-nucleon terms). 
The explicit-breaking effect on the nucleon-mass can be incorporated by coupling the nucleon field to the spurion field $\chi$, in a chiral-invariant way, as in the meson sector. The leading order correction thus arises at ${\cal O}(p^2)$ to take the form 
\begin{align} 
{\cal L}_{(2)}^{m_N} = - b_0 \bar{\hat{N}} (\hat{\chi} + \hat{\chi}^\dag) \hat{N}
- b_1 {\rm tr}[\hat{\chi} + \hat{\chi}^\dag] \bar{\hat{N}} \hat{N} 
\,. \label{Lag:mN}
\end{align}  
Expanding the fields in the sum of ${\cal L}_{(1)}^N$ and ${\cal L}_{(2)}^{m_N}$ 
leads to the nucleon masses and its splitting at the leading order 
of the chiral perturbation: 
\begin{align}
m_n &= m_0 + 2 B_0 [ (b_0 + 2 b_1) (m_u + m_d) + b_0 (m_d - m_u)]  \,, \notag \\ 
m_p &= m_0 + 2 B_0 [(b_0 + 2 b_1) (m_u + m_d) - b_0 (m_d - m_u)] \,, \notag \\ 
\Delta m_N & \equiv m_n - m_p = 4 B_0 b_0 (m_d - m_u)
\,. \label{delta-mN}
\end{align}

\subsection{Coupling scalaron to baryon chiral perturbation theory at leading order}

Now we make the chameleon, the scalaron field $\varphi$, coupled to 
the BChPT, which is at the leading order of derivative 
expansion (up to ${\cal O}(p^2)$) 
described by ${\cal L}_{(2)} + {\cal L}_\chi + {\cal L}_{(1)}^N + {\cal L}_{(2)}^{m_N}$ 
in Eqs.~\eqref{Lag:p2}, \eqref{Lag:chi}, \eqref{Lag:N} and \eqref{Lag:mN}.  
To this end, only we need to do is just transform the Minkowski space-time metric $g_{\mu\nu}$
in Jordan frame into the one in Einstein frame, $g_{\mu\nu} \to 
\tilde{g}_{\mu\nu} = e^{2\sigma} g_{\mu\nu}$ by the Weyl transformation, 
where $\sigma = \kappa \varphi/\sqrt{6}$ with $\kappa$ being inverse of reduced Planck mass ($1/M_{\rm pl}=1/\sqrt{8 \pi G}$). 
Then we can find coupling terms between the scalaron field $\varphi$ and 
the nucleon-pion system, in the same manner as done in~\cite{Katsuragawa:2016yir} 
for the standard-model sector. 
After canonically normalizing the nucleon field by transforming $\hat{N}$ as 
$\hat{N} \to e^{3/2 \sigma} \hat{N}$, we find the action corresponding to the nucleon part in Eqs.~\eqref{Lag:N} and \eqref{Lag:mN}: 
\begin{align} 
 S_{{\rm BChPT}{\cal O}(p^2)} [\sigma]
& = \int d^4 x \sqrt{-\tilde{g}} 
 \Bigg[ 
 \bar{\hat{N}} i \gamma^\mu (\partial_\mu - i \alpha_\mu^{||} ) \hat{N} 
+ g_A \bar{\hat{N}} i \gamma^\mu \gamma^5 \alpha_\mu^\perp \hat{N} 
\notag \\ 
& 
- e^{-\sigma} \left( m_0 \bar{\hat{N}} \hat{N} 
+ b_0 \bar{\hat{N}} (\hat{\chi} + \hat{\chi}^\dag) \hat{N}
+ b_1 {\rm tr}[\hat{\chi} + \hat{\chi}^\dag] \bar{\hat{N}} \hat{N} 
 \right)
  \Bigg] 
  \,. \label{varphi-BCHPT}
\end{align}
Note that
this scale transformation gives rise to the scale-anomaly at the quantum level of 
the baryon chiral perturbation theory, just like in the standard-model case coupled to 
the scalaron discussed in~\cite{Katsuragawa:2016yir}. This anomaly can be 
matched to the one induced from the scalaron couplings to 
the quarks and leptons in the SM, to reproduce precisely the same 
as the scalaron couplings to diphoton and digluon, derived in the reference.

From the action in Eq.~\eqref{varphi-BCHPT} we can easily read off the nucleon mass formulae similar to those displayed in 
Eq.~\eqref{delta-mN}, but modified by the presence of the scalaron profile as 
\begin{align}
m_n(\sigma) &= \left( m_0 + 2 B_0 [ (b_0 + 2 b_1) (m_u + m_d) + b_0 (m_d - m_u)] \right)e^{-\sigma} = m_n \cdot e^{-\sigma}
 \,, \notag \\ 
m_p(\sigma) &= \left( m_0 + 2 B_0 [(b_0 + 2 b_1) (m_u + m_d) - b_0 (m_d - m_u)] 
\right) e^{-\sigma} 
= m_p \cdot e^{-\sigma}
\,, \notag \\ 
\Delta m_N(\sigma) & = \left[ 
4 B_0 b_0 (m_d - m_u) \right] e^{-\sigma} 
= \Delta m_N \cdot e^{-\sigma}
\,. \label{delta-mN-varphi}
\end{align} 
It is interesting to note that at the leading order of the BChPT, the scalaron effects on the nucleon mass 
by coincidence  looks just like the Brown-Rho (BR) scaling~\cite{Brown:1991kk}, or the leading-order scale symmetry relation~\cite{Li:2016uzn,Li:2017hqe} 
as a sort of the extension of the BR scaling.

\bibliographystyle{JHEP}
\bibliography{References}

\end{document}